\documentclass[useAMS,usenatbib]{mn2e}
\usepackage{graphicx}
\usepackage{txfonts}
\usepackage{hyperref}
\usepackage[usenames, dvipsnames]{color}
\usepackage{enumitem}
\def\aj{AJ}%
\def\actaa{Acta Astron.}%
\def\araa{ARA\&A}%
\def\apj{ApJ}%
\def\apjl{ApJ}%
\def\apjs{ApJS}%
\def\aap{A\&A}%
\def\mnras{MNRAS}%
\def\pasp{PASP}%
\def\procspie{Proc.~SPIE}%

\usepackage[utf8]{inputenc}
\usepackage{pdflscape}

\hypersetup{draft} 

\newcommand{\sqdeg}{\mbox{deg$^{2}$}}

\newcommand{\jks}{\mbox{$J\!-\!K_{\rm s}$}}
\newcommand{\yks}{\mbox{$Y\!-\!K_{\rm s}$}}
\newcommand{\ks}{\mbox{$K_{\rm s}$}}

\newcommand{\dmo}{\mbox{$(m\!-\!M)_{0}$}}

\newcommand{\av}{\mbox{$A_V$}}
\newcommand{\avjk}{\mbox{$A_V^{JK_\mathrm{s}}$}}
\newcommand{\avyk}{\mbox{$A_V^{YK_\mathrm{s}}$}}

\newcommand{\feh}{\mbox{\rm [{\rm Fe}/{\rm H}]}}
\newcommand{\mh}{\mbox{\rm [{\rm M}/{\rm H}]}}
\newcommand{\MH}{\mbox{\rm [{\rm M}/{\rm H}]}}
\newcommand{\Msun}{\mbox{$\mathrm{M}_{\odot}$}}

\newcommand{\chisq}{\mbox{$\chi^2$}}
\newcommand{\chisqmin}{\mbox{$\chi^2_{\rm min}$}}

\title[SFH of the SMC]{The VMC survey -- XXXI. The spatially resolved star formation history of the main body of the Small Magellanic Cloud}

\author[Rubele et al.]{Stefano Rubele$^{1,2}$, 
Giada Pastorelli$^{1}$,   
L\'eo Girardi$^{2}$, 
Maria-Rosa L. Cioni$^{3}$, \newauthor 
Simone Zaggia$^{2}$, 
Paola Marigo$^{1}$, 
Kenji Bekki$^{4}$, 
Alessandro Bressan$^{5}$,\newauthor  
Gisella Clementini$^{6}$,
Richard de Grijs$^{7,8,9}$,
Jim Emerson$^{10}$,
Martin A.T. Groenewegen$^{11}$,\newauthor 
Valentin D. Ivanov$^{12,13}$, 
Tatiana Muraveva$^{6}$,
Ambra Nanni$^{1}$,
Joana M. Oliveira$^{14}$, \newauthor  
Vincenzo Ripepi$^{15}$, 
Ning-Chen Sun$^{8,16}$, 
Jacco Th.\ van Loon$^{14}$
\\
$^{1}$ Dipartimento di Fisica e Astronomia, Universit\`a di Padova, Vicolo dell’Osservatorio 2, I-35122 Padova, Italy\\
$^2$ Osservatorio Astronomico di Padova -- INAF, Vicolo dell'Osservatorio 5, I-35122 Padova, Italy\\
$^3$ Leibniz-Institut für Astrophysik Potsdam, An der Sternwarte 16, D-14482 Potsdam, Germany\\ 
$^{4}$ CRAR, M468, University of Western Australia, 35 Stirling Hwy, 6009 Crawley, Western Australia, Australia\\
$^5$ SISSA, via Bonomea 265, I-34136 Trieste, Italy \\
$^{6}$ INAF -- Osservatorio di Astrofisica e Scienza dello Spazio di Bologna, via Piero Gobetti 93/3, 40129, Bologna, Italy \\
$^7$ Department of Physics and Astronomy, Macquarie University, Balaclava Road, North Ryde, NSW 2109, Australia\\
$^8$ Kavli Institute for Astronomy and Astrophysics, Peking University, Yi He Yuan Lu 5, Hai Dian District, Beijing 100871, China\\
$^9$ International Space Science Institute--Beijing, 1 Nanertiao, Zhongguancun, Hai Dian District, Beijing 100190, China\\
$^{10}$ Astronomy Unit, School of Physics and Astronomy, Queen Mary University of London, Mile End Road, London E1 4NS, UK\\
$^{11}$ Koninklijke Sterrenwacht van Belgi\"e, Ringlaan 3, B--1180 Brussels, Belgium\\
$^{12}$ European Southern Observatory, Ave. Alonso de Cordova 3107, Vitacura, Santiago, Chile\\
$^{13}$ European Southern Observatory, Karl-Schwarzschild-Str. 2, D-85748 Garching bei München, Germany\\
$^{14}$ Lennard-Jones Laboratories, Keele University, ST5 5BG, UK\\
$^{15}$ INAF -- Osservatorio Astronomico di Capodimonte, via Moiariello 16, 80131, Naples, Italy\\
$^{16}$ Department of Astronomy, Peking University, Yi He Yuan Lu 5, Hai Dian District, Beijing 100871, China
 }

\date{To appear on MNRAS, May 11, 2018}

\begin{document}

\maketitle
\label{firstpage}

\begin{abstract}
We recover the spatially resolved star formation history across the entire main body and Wing of the Small Magellanic Cloud (SMC), using fourteen deep tile images from the VISTA survey of the Magellanic Clouds (VMC), in the $YJ\ks$ filters. The analysis is performed on 168 subregions of size 0.143~deg$^2$, covering a total contiguous area of $23.57$~\sqdeg. We apply a colour--magnitude diagram (CMD) reconstruction method that returns the best-fitting star formation rate SFR$(t)$, age--metallicity relation, distance and mean reddening, together with their confidence intervals, for each subregion. With respect to previous analyses, we use a far larger set of VMC data, updated stellar models, and fit the two available CMDs ($\yks$ versus $\ks$ and $\jks$ versus $\ks$) independently. The results allow us to derive a more complete and more reliable picture of how the mean distances, extinction values, star formation rate, and metallicities vary across the SMC, and provide a better description of the populations that form its Bar and Wing. We conclude that the SMC has formed a total mass of $(5.31\pm0.05)\times10^8$~\Msun\ in stars over its lifetime. About two thirds of this mass is expected to be still locked in stars and stellar remnants. 50 per cent of the mass was formed prior to an age of 6.3 Gyr, and 80 per cent was formed between 8 and 3.5~Gyr ago. 
We also illustrate the likely distribution of stellar ages and metallicities in different parts of the CMD, to aid the interpretation of data from future astrometric and spectroscopic surveys of the SMC.
\end{abstract}


\section{Introduction}
\label{sec:intro}
 
The Magellanic Clouds represent the best possible galaxies for the derivation of their spatially resolved star formation histories (SFH). They are not only close enough to be entirely resolved into stars with ground-based telescopes, even down to the depth of the oldest main sequence turn-offs, but also are just moderately affected by interstellar extinction and foreground Milky Way stars. While the Large Magellanic Cloud (LMC) presents a relatively simple disc+bar structure, the SMC is known to present a more complex geometry, with indications of significant depths along several lines of sight \citep[][]{gardiner91,degrijs15,scowcroft16,ripepi17,Muraveva18}, and evidence of two different structures along its eastern Wing \citep{Nidever_etal11, piatti15, subramanian17}. As shown by \citet[][]{HZ01}, such depths do not hamper the quantitative derivation of the SFHs via colour--magnitude diagram (CMD) reconstruction methods.

Previous literature about the SFH of the SMC is dominated by studies based on deep optical photometry. Among these, no previous work rivals the ample SMC area ($18$~deg$^2$) covered by the Magellanic Clouds Photometric Survey (MCPS; \citealt{zari02}), which was analysed by \citet{HZ04}. These authors reached important conclusions about the global SFH of the SMC, indicating for instance that 50 per cent of its stellar mass formed at ages prior to 8.4~Gyr ago, the presence of enhanced star formation at ages of 2.5, 0.4 and 0.06~Gyr, and the presence of a large ring-like structure in the 2.5~Gyr burst. Many other works are dedicated to the analysis of deep optical photometry of selected areas, using either dedicated ground-based surveys \citep[e.g.][]{noel07,noel09} or the {\it Hubble Space Telescope} (HST) \citep[e.g.][]{cignoni12,cignoni13,weisz13}. These works generally confirm a wide variation in the SFH from field to field \citep[e.g.][]{cignoni13}, at least in the central SMC regions. Several small-area studies appear to confirm the few periods of enhanced star formation claimed by \citet{HZ04}, although they are usually found at slightly different ages. In regions more distant than about $2.7\degr$ from the SMC centre, the SFH appears to be much more uniform, and the surface brightness decays exponentially \citep[][]{noel07}. This simple picture of the SMC outskirts is challenged in the so-called SMC Wing, which shows signs of recent star formation stretching to larger radii \citep{irwin90}, and in the Magellanic Bridge, which shows stellar overdensities attributable either to tidal interactions between the two Magellanic Clouds, or to an overlap between their halos \citep[][]{skowron14}. 

A few works aimed to constrain the SFH of the SMC from the analysis of wide-area near-infrared surveys \citep[e.g.][]{cioni06,reza14}. Compared to works based on deep optical data, they rely on smaller numbers of stars, mostly located in the asymptotic giant branch (AGB) phase. Therefore, they are more affected by small-number statistics and by the significant uncertainties of theoretical models of evolved stars. Nonetheless, they have provided independent evidences of past periods of enhanced star formation, as for instance those inferred at ages $\sim\!0.7$~Gyr and $\sim\!6$~Gyr by \citet{reza14}.

The VISTA survey of the Magellanic Clouds \citep[VMC;][]{cioni11} represents a major effort to provide deep and homogeneous near-infrared photometry across the Magellanic Clouds, so that their SFHs and basic geometry can be derived with minimal interference owing to the effects of interstellar dust. The VMC is an ESO public survey using the VIRCAM camera of the VISTA 4-m telescope \citep{vista} in the $Y$, $J$, and $\ks$ filters. The survey has been designed so that its photometry reaches the turn-off region of the oldest ($\sim\!13$~Gyr) stellar populations in the Magellanic Clouds, even in the most crowded regions of the LMC bar \citep[see][]{kerber09}. SFHs were already derived for a few regions of the LMC by \citet{Rubele_etal12}, and for a large non contiguous section of the SMC by \citet{Rub15}. In both cases, the data also allowed us to derive clear indications about the geometry of the regions observed.  Complementary information on the geometry was provided by the near-infrared properties of the variables, when using VMC data in combination with the OGLE and EROS2 surveys \citep[see][]{ripepi12,ripepi15,ripepi17, moretti14,moretti16,Muraveva18}.

Once derived, the SFHs can be useful for a variety of applications, from the exploration of the mechanisms that drive the star formation and chemical evolution in dwarf galaxies over long timescales, to the discussion of systematic effects in the magnitudes of stellar standard candles, to the calibration of stellar models (at least for fast evolutionary phases not involved in the SFH derivation). 

In this paper, we revisit the spatially resolved SFH of the SMC. This revision is motivated by: 
\begin{enumerate}
\item a further, significant increase in the area and depth covered by the VMC observations (as described in Sect.~\ref{sec:data}), which now reaches 100 per cent completion for the entire main body of the SMC, covering a contiguous area of $23.57$~deg$^2$ \citep[30 per cent larger than the one analysed by][]{HZ04};  
\item a few significant improvements in our analysis, regarding the photometric zeropoints and stellar models (Sect.~\ref{sec:sfh}). 
\end{enumerate}
These novelties are significant enough to motivate a renewed discussion of the SMC results. They also allow us to derive more accurate global quantities, such as the total mass of stars formed, which was not possible in earlier analyses based on smaller data sets. These improved results are described and discussed in Sects.~\ref{sec:D_AV} to \ref{sec:conclu}. Furthermore, the derived SFH, extinction and distance values are at the basis of other population and stellar evolution work being carried out with the aid of additional SMC data (Pastorelli et al., in prep.). 
 
Finally, we note that \citet{sun08} analyse the SMC's young star formation, using essentially the same data but very different methods. That paper uses the detailed spatial resolution available in VMC data to identify young stellar structures and characterize their size and mass distributions. In the present work, instead, we aim at deriving the SFH, following a method which requires the data being grouped into spatial bins. These spatial bins are certainly larger than the resolution adopted by \citet{sun08}, but still small enough to allow us to discuss, in a quantitative way, the spatial distribution of the populations of all ages. Therefore, both works provide complementary (and overall consistent) views of the SMC stellar populations, at different spatial scales and age ranges.

\begin{table*}
  \caption{VMC tiles used in this work.}
\label{tab_tiles}
\begin{tabular}{cccccccccc}
\hline
\hline
Tile & \multicolumn{2}{c}{Central coordinates} & completeness & error [mag] & $50\%$ completeness  & Comments\\
     & $\alpha$ (h:m:s, J2000) &  $\delta$ (deg:m:s, J2000) &  \multicolumn{2}{c}{at \ks=20.45 mag} & mag.    &         \\
\hline
SMC 3\_2  &  00:23:35.544 & $-$74:06:57.240 &  0.87  & 0.184 & 20.95  & SW extreme of Bar\\
SMC 3\_3  &  00:44:55.896 & $-$74:12:42.120 &  0.77  & 0.212 & 20.89 & S extreme of Bar\\
SMC 3\_4  &  01:06:21.120 & $-$74:10:38.640 &  0.87  & 0.196 & 20.97  & S part of Wing\\   
SMC 3\_5  &  01:27:30.816 & $-$74:00:49.320 &  0.91  & 0.170 & 20.92 & SE part of Wing\\   
SMC 4\_2  &  00:25:14.088 & $-$73:01:47.640 &  0.90  & 0.165 & 21.28 & central extreme part of Bar\\   
SMC 4\_3  &  00:45:14.688 & $-$73:07:11.280 &  0.51  & 0.378 & 20.51 & SW part of densest Bar \\   
SMC 4\_4  &  01:05:19.272 & $-$73:05:15.360 &  0.75  & 0.245 & 21.09 & central Bar, slightly towards Wing\\    
SMC 4\_5  &  01:25:11.016 & $-$72:56:02.040 &  0.92  & 0.156 & 21.22 & central part of Wing\\    
SMC 5\_3  &  00:44:49.032 & $-$72:01:36.120 &  0.82  & 0.211 & 20.96 & NW of densest Bar\\    
SMC 5\_4  &  01:04:26.112 & $-$71:59:51.000 &  0.82  & 0.223 & 21.08 & NE part of densest Bar\\     
SMC 5\_5  &  01;23:04.944 & $-$71:51:47.880 &  0.91  & 0.192 & 21.13 & N part of Wing\\     
SMC 6\_3  &  00:45:48.768 & $-$70:56:08.160 &  0.91  & 0.184 & 21.10 & $\sim1.5$~deg NW of main body\\
SMC 6\_4  &  01:03:49.944 & $-$70:53:34.440 &  0.87  & 0.163 & 20.98 & $\sim1.7$~deg N of main body\\   
SMC 6\_5  &  01:21:22.488 & $-$70:46:10.920 &  0.94  & 0.156 & 21.15 & $\sim3$~deg NE of main body\\    
\hline
\end{tabular}
\end{table*}

\begin{figure*}
\resizebox{0.9\hsize}{!}{\includegraphics{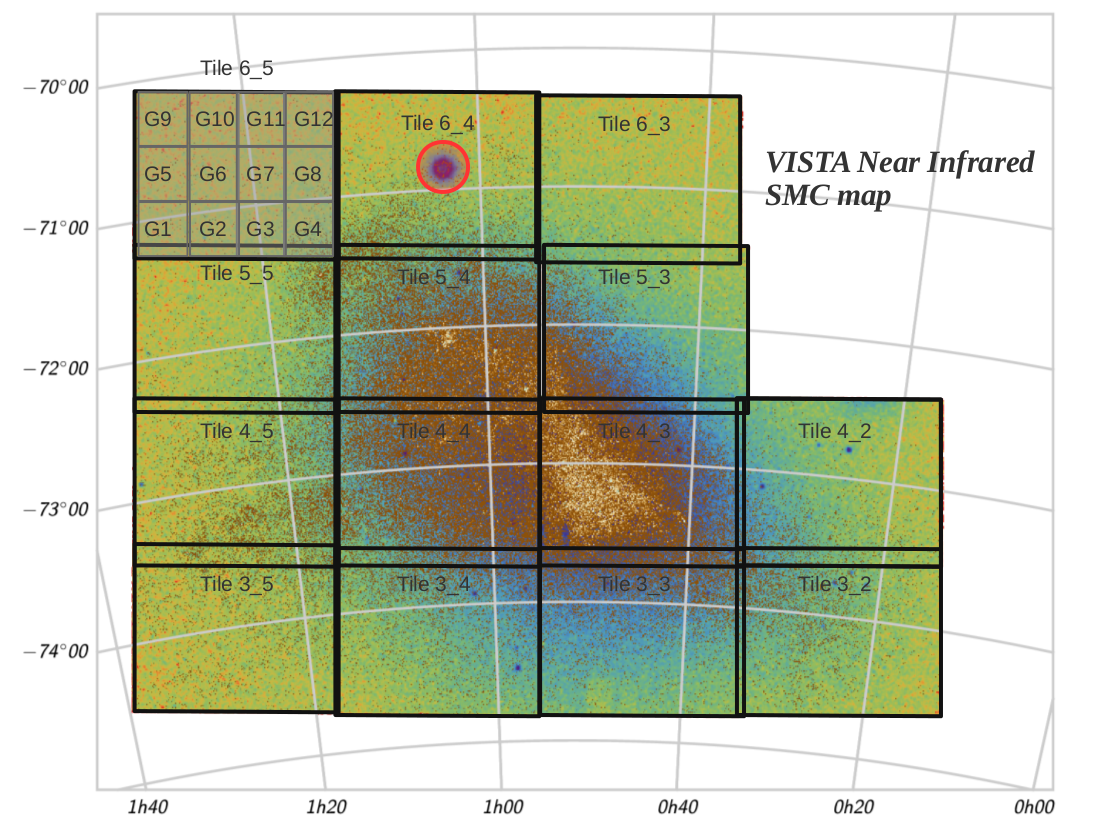}}
\caption{Stellar density distribution of the SMC as obtained from the VMC data. The black rectangles delimit the deep tiles used in our analysis, as listed in Table~\ref{tab_tiles}. The smaller grey rectangles overplotted on tile SMC 6\_5 illustrate how the tiles are divided into twelve subregions. For our subsequent analysis, an area of $254$~arcmin$^2$ has been removed from the data around the Milky Way star cluster NGC~362 in tile SMC 6\_4, as indicated by the red circle.}
\label{fig:SMCmap}
\end{figure*}

\section{Data and photometry}
\label{sec:data}

\subsection{The VMC data}
\label{sec:selecteddata}
 
We refer to \citet{cioni11} for a general description of the VMC survey, and to \citet{Rub15} for a more detailed discussion and illustration of the properties of the SMC data. Suffice it to recall that the SMC galaxy is covered by 27 VISTA tiles, each one covering about $1.77$~deg$^2$ on the sky, and extending up to $\sim3.5\degr$ from the SMC centre. In this work, we investigate the fourteen central SMC tiles listed in Table~\ref{tab_tiles}, which cover the main bar-like feature seen in projection in the SMC (hereafter the ``Bar''; Fig.~\ref{fig:SMCmap}) -- with the only exception of a narrow $0.145\degr\times1\degr$ gap between tiles SMC 5\_3 and 5\_4\footnote{The gap covers just 0.6 per cent of the analysed area, hence it does not affect our results in a significant way.} -- and the SMC inner Wing, for a contiguous area of 23.57~deg$^2$. All these tiles have 100 per cent completion in the \ks\ band, which correspond to at least $12$ epochs and at least $9000$~sec of integration time.

The background image in Fig.~\ref{fig:SMCmap} is a density map of all VMC sources with $\ks\!<\!18$~mag and $\ks$ errors smaller than $0.2$~mag. Since this magnitude cut includes the red clump (RC) and the upper part of the red giant branch (RGB), the map is dominated by intermediate-age and old stellar populations. The superimposed brown-white density points code the distribution of young stellar populations selected from the colour cut $\yks<0.5$~mag. Central SMC  regions are are well observed without any limitation due to confusion and crowding.

Also note that we decided to ignore tile SMC 5\_2 in this work, because it is dominated by the 47~Tuc globular cluster \citep[see][]{li14,zhang15,cioni16,nieder18,sun08}. The presence of 47~Tuc is also apparent as a small stellar overdensity in the NW section of tile SMC 4\_2. Moreover, tile SMC 6\_4 contains the compact Milky Way globular cluster NGC~362, which dominates the star counts in the CMD of the two sub regions G6 and G7 (see top left of Fig.~\ref{fig:SMCmap}). We remove the latter object by applying a cut of radius $9 \arcmin$ from the cluster centre, located at $\mathrm{RA} =15.809^\circ$ and $\mathrm{Dec}=-70.8489^\circ$.  

\begin{figure*}
\resizebox{0.45\hsize}{!}{\includegraphics{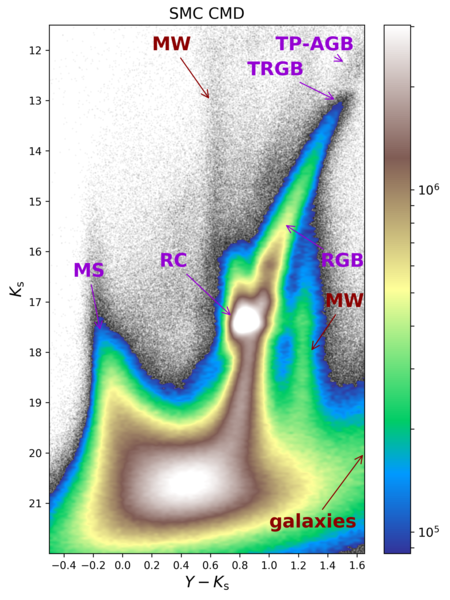}}
\resizebox{0.45\hsize}{!}{\includegraphics{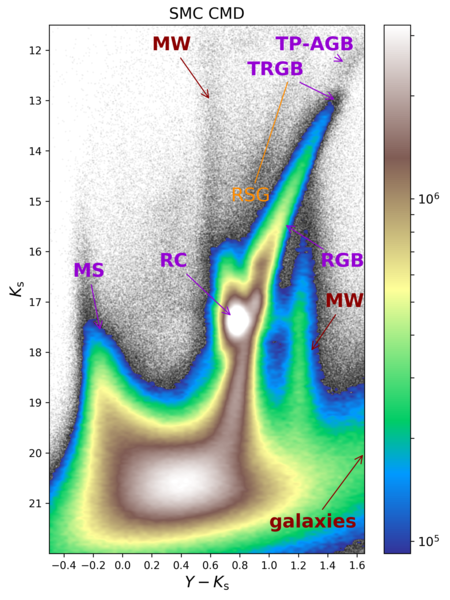}}
\\
\resizebox{0.45\hsize}{!}{\includegraphics{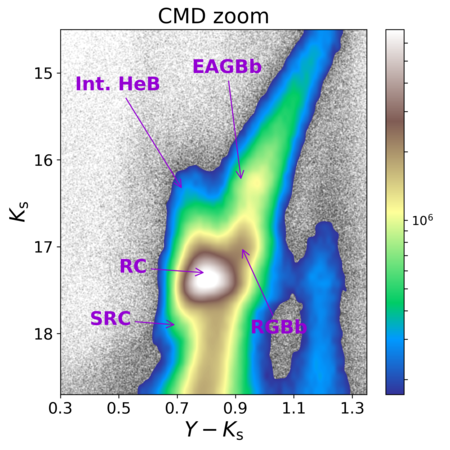}}
\resizebox{0.45\hsize}{!}{\includegraphics{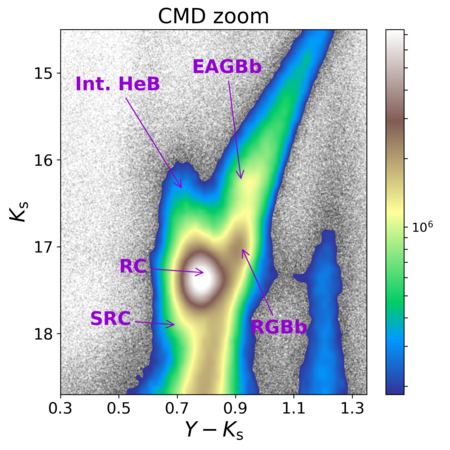}}
%
\caption{\ks\ versus \yks\ CMDs for the entire SMC area analysed in this work (see Fig.~\ref{fig:SMCmap}). The \textbf{left-hand panels} show the original PSF photometry, while the \textbf{right-hand panels} show the same after the data for every subregion has been corrected to the same reference value of distance modulus and extinction, namely $\dmo=18.9$~mag and $\av=0.35$~mag. The \textbf{top panels} show the entire CMD region relevant for this work, while the \textbf{bottom panels} zoom into the feature-rich region around the red clump (RC).  The arrows point to some of the most prominent CMD features: In the SMC (magenta arrows and labels), we have the main sequence (MS), the red giant branch (RGB), the red clump (RC); at its brightest and coolest extremity, there is a well-defined tip of the RGB (TRGB), a population of thermally pulsing asymptotic giant branch (TP-AGB) stars, and a well-defined strip of He-burning stars that we refer to as red supergiants (RSG). Features around the RC include: the secondary red clump (SRC), intermediate-mass core-He burning stars (Int.~HeB), the RGB bump (RGBb), and the early asymptotic giant branch bump (EAGBb).  In addition, we can clearly see the foreground/background populations indicated by the red arrows: the presence of very faint and red background galaxies (most of which are actually redder than the limits shown), and two long nearly vertical features corresponding to foreground Milky Way stars, the bluest at $\yks\simeq0.6$~mag caused by the turn-off of populations of intermediate to old ages, and the reddest at $\yks\simeq1.25$~mag caused by the ``CMD kink'' of low-mass, cool M dwarfs. See also \citet{sun08} for a complementary view, that better discusses the features caused by the young SMC populations.
}
\label{fig:CMDtotal}
\end{figure*}

\subsection{Image mosaicking, photometry and artificial star tests}
\label{sec:psfphotometry}
 
We use v1.3 of the VMC data retrieved from the VISTA Science Archive \citep[VSA;][]{Hambly_etal04}\footnote{\url{http://horus.roe.ac.uk/vsa/}}. 
Our data analysis starts from the pawprint images, already processed and calibrated by the VISTA Data Flow System \citep[VDFS;][]{Emerson_etal04, Irwin_etal04} pipeline. We homogenised individual pawprints point spread function (PSF), and then combined them into deep tile images on which we performed the PSF photometry. 
Subsequently we correlate the photometry in the three bands ($YJ\ks$) using a $1\arcsec$ matching radius to generate a multi-band catalogue. Finally, we apply the aperture correction using as reference the VSA data release v1.3 (see \citealt{Cross_etal12} and \citealt{Irwin_etal04} for details). For a detailed description of the methodology see \citet{Rub15}. Fig.~\ref{fig:CMDtotal} gives an idea of the overall quality of the entire data set, based on the \ks\ versus \yks\ CMD.

A large number of artificial star tests (ASTs) were performed on tile images, so as to map the distributions of photometric errors and completeness, as a function of colour, magnitude, and position. The process is the same as that extensively described and illustrated by \citet{Rubele_etal12,Rub15}. 
In all our tiles the 50 per cent completeness limits correspond to magnitudes fainter than the magnitude cut applied in the subsequent analysis (which are 21.25, 20.95 and 20.45~mag in the $Y$, $J$ and $\ks$ filters, respectively). The completeness at $\ks=20.45$~mag, averaged for each tile, is presented in Table~\ref{tab_tiles}. 

\section{The SFH recovery}
\label{sec:sfh}

\subsection{The method}
\label{sec:method}
As in \citet{Rub15}, the derivation of the SFH simply consists of finding the linear combination of partial models that best fits the observed Hess diagrams, that is, the stellar density in the CMDs. The partial models themselves are the theoretical realisations of simple stellar populations, with a known total mass of formed stars, fixed values for the true distance modulus \dmo\ and extinction \av, and covering small ranges in age and initial metallicity. Partial models also  incorporate a simulation of the photometric errors and incompleteness distributions derived from the ASTs. In addition, there is a partial model representing the foreground Milky Way population, derived from the latest version of the TRILEGAL code \citep{Girardi_etal05,Girardi_etal12}. The best-fitting solution is found by application of the StarFISH optimization code of \citet{HZ01}, and its fitting coefficients are directly translated into a SFH. Subsequent searches are done to locate the \dmo\ and \av\ values that minimise the model-data $\chi^2$ -- hence identifying $\chisqmin$ -- and to provide the confidence levels of all best-fitting parameters. 

In this work, every tile is divided into twelve subregions of areas equal to 0.143~deg$^2$, as illustrated for the tile SMC 6\_5 in Fig.~\ref{fig:SMCmap}. This subregion size represents the minimum area (and star counts) necessary to recover the young SFH with random errors smaller than $\sim\!10$ per cent in the central SMC tiles, and yet it allows us to achieve a similar accuracy for the old SFH in the most external tiles. A complete discussion of how the SFH errors scale with the subregion area and stellar density, for populations of different age, can be found in \citet{kerber09}.

With this general procedure in mind, we now describe the few changes with respect to our previous analysis of the SMC using VMC data.

\subsection{Changes in the partial models}
\label{sec:partialmodels}

\begin{table}
\caption{Grid of SMC stellar partial models used in the SFH recovery.}
\label{tab_amr}
\centering
\begin{tabular}{c|ccccc}
\hline\hline
$\log(t/{\rm yr})$ & $\mh_1$ & $\mh_2$ & $\mh_3$ & $\mh_4$ & $\mh_5$ \\
 & dex & dex & dex & dex & dex \\
\hline
6.9 & $-$0.40 & $-$0.55 & $-$0.70 & $-$0.85 & $-$1.00 \\
7.4 & $-$0.40 & $-$0.55 & $-$0.70 & $-$0.85 & $-$1.00 \\
7.8 & $-$0.40 & $-$0.55 & $-$0.70 & $-$0.85 & $-$1.00 \\
8.1 & $-$0.40 & $-$0.55 & $-$0.70 & $-$0.85 & $-$1.00 \\
8.3 & $-$0.40 & $-$0.55 & $-$0.70 & $-$0.85 & $-$1.00 \\
8.5 & $-$0.40 & $-$0.55 & $-$0.70 & $-$0.85 & $-$1.00 \\
8.7 & $-$0.40 & $-$0.55 & $-$0.70 & $-$0.85 & $-$1.00 \\
8.9 & $-$0.40 & $-$0.55 & $-$0.70 & $-$0.85 & $-$1.00 \\
9.1 & $-$0.55 & $-$0.70 & $-$0.85 & $-$1.00 & $-$1.15 \\
9.3 & $-$0.55 & $-$0.70 & $-$0.85 & $-$1.00 & $-$1.15 \\
9.5 & $-$0.70 & $-$0.85 & $-$1.00 & $-$1.15 & $-$1.30 \\
9.7 & $-$0.85 & $-$1.00 & $-$1.15 & $-$1.30 & $-$1.45 \\
9.9 & $-$1.15 & $-$1.30 & $-$1.45 & $-$1.60 & $-$1.75 \\
10.075 & $-$1.45 & $-$1.60 & $-$1.75 & $-$1.90 & $-$2.05 \\
\hline
\end{tabular}
\end{table}

There are essentially three changes in the definition of partial models, with respect to \citet{Rub15}:

\paragraph*{Updated evolutionary tracks and isochrones:} Partial models for this work have been derived from PARSEC v1.2S evolutionary tracks and isochrones \citep{bressan12, bressan15}\footnote{\url{http://stev.oapd.inaf.it/cmd}}. They represent a major revision of the previous \citet{Marigo_etal08} models used by \citet{Rubele_etal12}, and a moderate update of the PARSEC v1.1 models used by \citet{Rub15}. Regarding the latter, the most relevant changes are in (1) revised surface boundary conditions used in low-mass dwarfs \citep[see][]{chen14}\footnote{Low-mass dwarfs are not relevant to the modelling of the SMC populations, but are critical in the description of the foreground Milky Way stars.}, and (2) a large extension in the grid of initial masses and metallicities used to generate the evolutionary tracks and isochrones. Moreover, isochrones are now built with a revised algorithm \citep[available since][]{Marigo_etal17} which ensures a more reliable interpolation of all evolutionary features as a function of age (or initial mass) and metallicity \mh. The stellar models assume scaled-solar abundances of metals, so that $\mh\equiv\feh$.
 
\begin{figure*}
\resizebox{0.9\hsize}{!}{\includegraphics{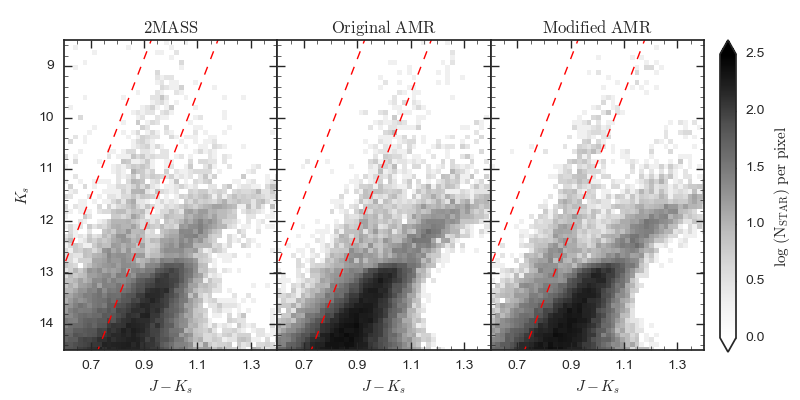}}
\caption{Comparison of the observed and simulated $\jks$ vs. $\ks$ Hess diagrams for the four central tiles (SMC 4\_3, SMC 4\_4, SMC 5\_3, and SMC 5\_4), at bright \ks\ magnitudes. Left-hand panel: 2MASS data. Middle panel: simulation computed with the AMR from \citet{Rub15}. Right-hand panel: simulation with the presently derived AMR (Sect.~\ref{sec:metal}), which is constrained to lower metallicities for young populations. The red dashed lines show the photometric criteria used by \citet{boyer11} to separate the RSG population of the SMC from the foreground and the TP-AGB stars. The use of the present AMR results in improvements of both the colour and the slope of the RSG sequence. } 
\label{fig:metlim}
\end{figure*}

\paragraph*{Limits to the metallicity of young populations:}
The 70 partial models for the SMC stars are built assuming finite widths in age and metallicity. Fourteen age bins are defined, each one with five different metallicity values (plus the Milky Way model). Table~\ref{tab_amr} specifies the adopted mean values of $\log(t/\mathrm{yr})$ and \mh. Most age bins span a 0.2~dex interval (on a logarithmic scale) except for the youngest partial models which span either 0.6 or 0.4 dex, and the oldest age bin which spans 0.15~dex. All partial models span $\Delta\mh=0.15$~dex in metallicity, distributed around an age--metallicity relation (AMR) in which older models are more metal poor. 
With respect to \citet{Rub15}, partial models for $\log(t/{\rm yr})<8.2$ were shifted by $-0.3$~dex, and those with  $8.2<\log(t/{\rm yr})<8.8$ by $-0.2$~dex; with these new limits, we limit the metallicities of the young populations in the SMC to $\mh$ values below $-0.325$~dex. This shift was adopted firstly to better comply with independent observations that indicate even lower metallicities for young SMC populations \citep[e.g.][]{hill99,davies15}, and secondly because the metallicities adopted by \citet{Rub15} were producing colours that were too red for the bright stars such as the red supergiants (RSGs). 

This latter problem is illustrated in Fig.~\ref{fig:metlim}, which compares the 2MASS data\footnote{For this example, we prefer to use 2MASS data in $J\ks$, because the VMC data is partially saturated at the bright magnitudes ($\ks\sim11$~mag) in which the RSG problem appears.} for the SMC (left-hand panel) with synthetic CMDs simulated using the mean AMR from \citet[][middle panel]{Rub15}, and that obtained in the present work (right-hand panel; see Sect.~\ref{sec:metal}) for the same areas but with the AMR constrained to lower metallicities at young ages. Both simulations use the same star formation rate as a function of age (see Sect.~\ref{sec:metal}). In the figure, the RSG population of the SMC appears between the red dashed lines, which were defined by \citet{boyer11} to separate them from the foreground Milky Way (to its left), and the TP-AGB stars (to its right). We can clearly notice that the use of the present AMR improves the colour of the young RSG sequence in the models: even if its colour still does not perfectly match the observed one from 2MASS, its slope in the CMD turns out to be correct now. Since the metallicity changes apply only to the very young populations, they do not affect the colour of the well-populated RGB (with ages $\log(t/\mathrm{yr})>9$), which appears nearly identical in the two simulations and in the 2MASS data. The metallicity change also does not affect the TP-AGB population in a significant way, since just a minor fraction of such stars have ages in the interval $8.0<\log(t/\mathrm{yr})<8.3$. We recall that the detailed counts and positions of TP-AGB stars in the plot depend on a lot of model details other than metallicity, as will be discussed on a forthcoming paper (Pastorelli et al., in prep.).

Having decided to set an upper limit to the metallicities of partial models, the question remains as to why the previous analysis by \citet{Rub15} favoured too metal-rich populations at young ages. The answer probably lies in the low sensitivity of the young main sequence turn-offs, in near-infrared filters, to metallicity. Indeed, the bulk of the young stars falling inside the colour-magnitude limits selected for our SFH analysis are in the main sequence. So, small errors in the extinction or in the model colours for these stars could have led the CMD reconstruction algorithm to favour  unlikely regions of AMR space, at least for these young stars. This does not happen for the red giants in the data -- which in general sample older populations -- because the RGB position and mean slope are very sensitive to the mean metallicity even at near-infrared colours.

\paragraph*{Changing the initial mass function (IMF):} In previous analyses of the SMC data we used the \citet{Chabrier03} log-normal IMF, which presents a steeper decrease in the number of massive stars than the canonical \citet{salpeter} IMF. Indeed, for stars with masses above 10~\Msun\ the \citet{Chabrier03} log-normal distribution translates into power-law slopes $>2$, well in excess of the $1.35$ and $1.30$ slopes of the \citet{salpeter} and \cite{Kroupa02} IMFs, respectively. The recent work by \citet{weisz15} suggests that a power-law IMF with a slope of $\sim1.3\pm0.1$ might better represent the stellar populations in the LMC -- while for young populations in M31 an excellent fit was found for an IMF slope of 1.45. Overall, these results suggest that the IMF for massive stars is significantly shallower than implied by the \citet{Chabrier03} log-normal IMF. Therefore, we decided to adopt the \cite{Kroupa02} IMF in this paper. The effect on the \chisq\ of our best-fitting models is modest, because (a) the fraction of the stellar counts coming from intermediate-mass and massive stars is very small ($\approx1$ per cent in the range $M>2$~\Msun), and (b) the cut in stellar mass (corresponding to the same limit in colour and depth in the CMDs) for our oldest partial model is at about $0.8$~\Msun. 
In this range of masses the two IMFs differ by just 6 per cent, with the \cite{Kroupa02} IMF predicting more low-mass stars than the \citet{Chabrier03} IMF.
 
\begin{figure}
\resizebox{0.99\hsize}{!}{\includegraphics{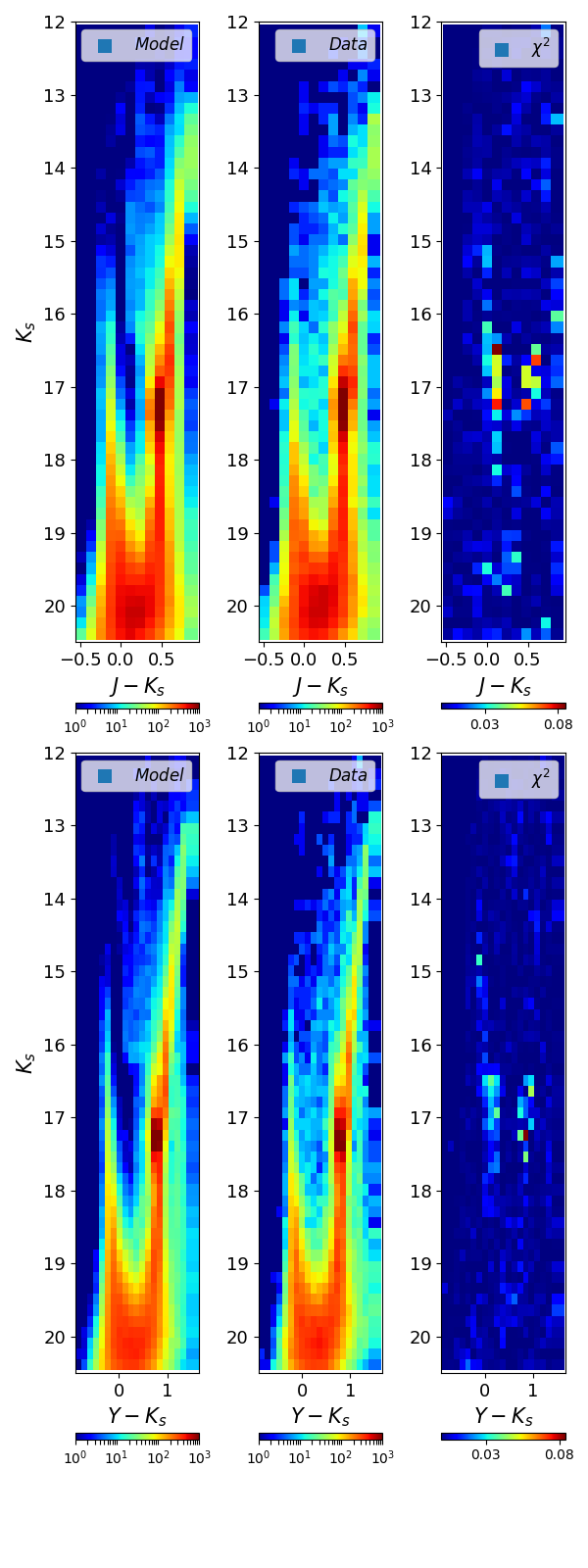}}
\caption{From left to right: Hess diagrams for the data, best-fitting model, and $\chi^2$, for  subregion G8 of tile SMC 5\_4. The top panels are for the \jks\ vs.\ \ks\ CMD, the bottom ones for \yks\ vs.\ \ks\ CMD. In the left-hand and middle panels, the colour scale indicates the number of stars per bin. 
}
\label{fig_sol}
\end{figure}

\begin{figure*}
\begin{minipage}{0.45\textwidth}
\resizebox{\hsize}{!}{\includegraphics{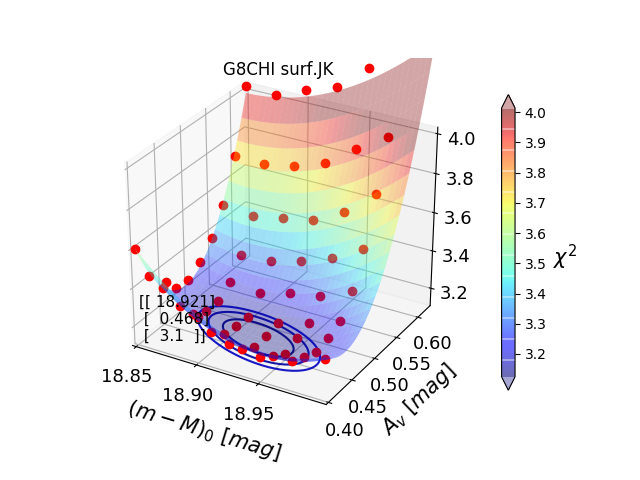}}
\resizebox{\hsize}{!}{\includegraphics{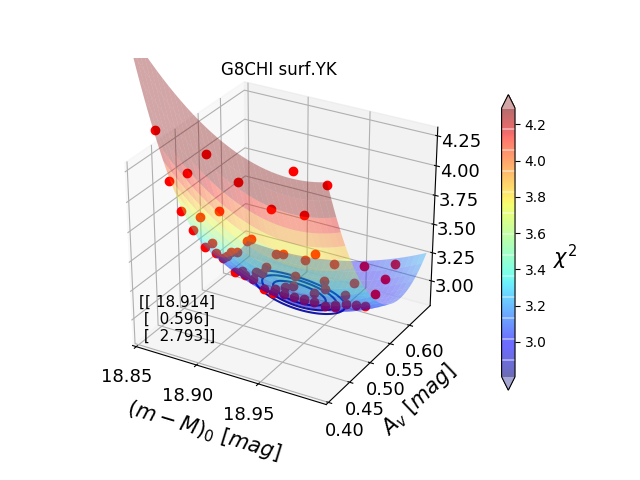}}
\end{minipage}
\hfill
\begin{minipage}{0.5\textwidth}
\resizebox{\hsize}{!}{\includegraphics{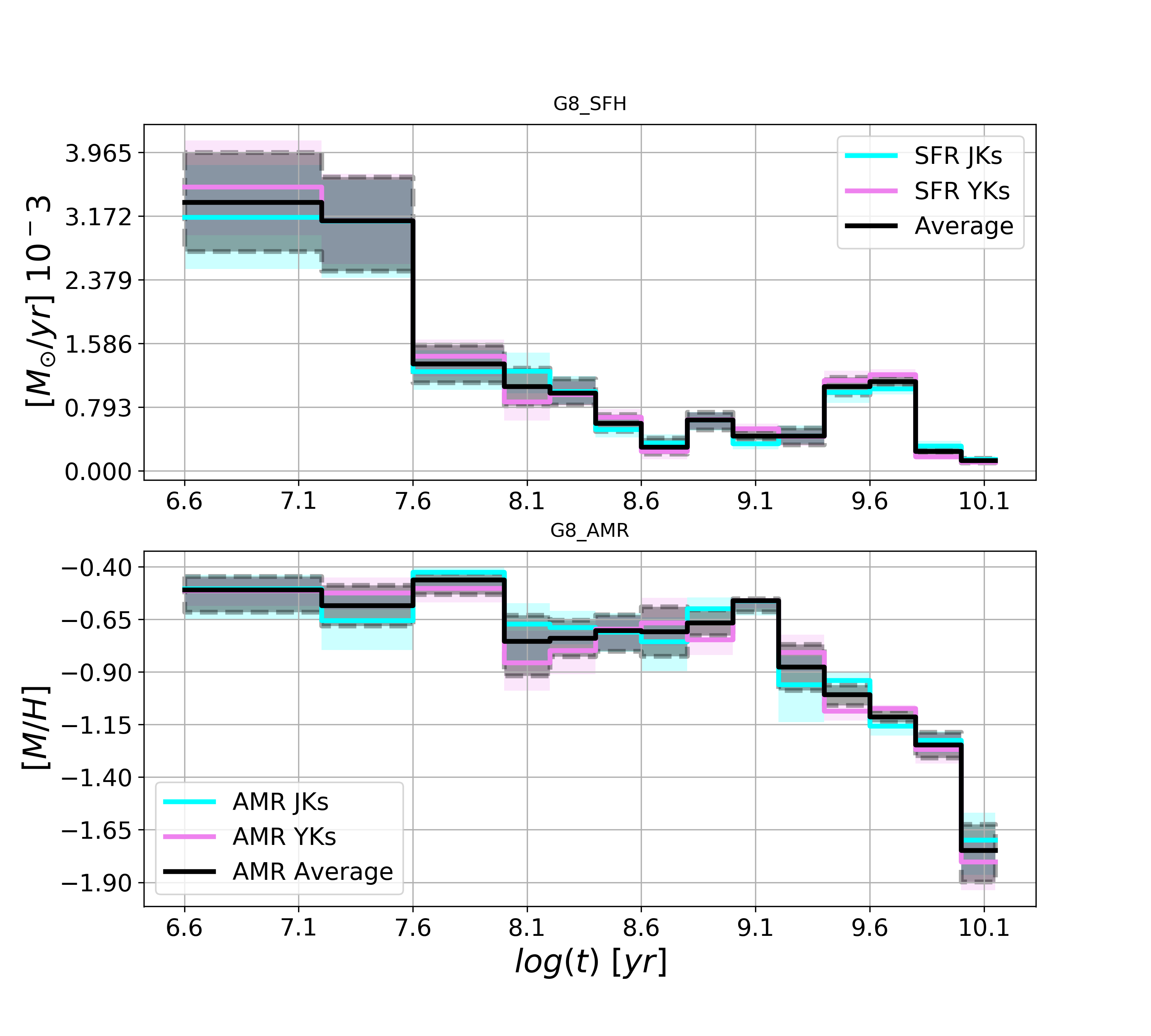}}
\end{minipage}
\caption{SFH results for subregion G8 of tile SMC 5\_4, which corresponds to a relatively dense region in the SMC Bar. The red dots in the {\bf left-hand panels} show the values of $\chi^2$ derived from StarFISH over a wide range of true distance moduli and extinctions, \dmo\ and \av, for both the \jks\ vs. \ks\ and \yks\ vs. \ks\ CMDs (top and bottom panels, respectively). The 3D surface is a simple second-order polynomial fit to this distribution, used to locate the best-fitting \dmo\ and \av\ and their confidence levels. The latter correspond to the three contour lines in the \dmo\ vs. \av\ plane, for the 1$\sigma$ (68 per cent),  $2\sigma$ (95 per cent) and $3\sigma$ (99.7 per cent) confidence levels. The {\bf right-hand panels} show the best-fitting solution in the form of the SFR$(t)$ in units of $\Msun\,{\rm yr}^{-1}$ versus the logarithm of age (top panel), and as the evolution of the mean metallicity (bottom panel); the solutions for the $J\ks$ and $Y\ks$ cases, and their averages, are marked as cyan, plum and black histograms, respectively. In both panels, the shaded areas show the random errors in the SFR$(t)$ and $\mh(t)$ relations -- again with cyan, plum and grey areas corresponding to the \jks, \yks\ cases, and their averages. 
}
\label{fig_sfr}
\end{figure*}

\subsection{Decoupling the two CMDs} 
\label{sec:decoupling}
\citet{Rub15} performed an analysis of the two available CMD/Hess-diagrams --  namely $\yks$ vs. $\ks$ and $\jks$ vs. $\ks$ -- simultaneously, using a common value for the extinction \av. This \av\ was then translated into $A_Y$, $A_J$ and $A_{K_\mathrm{s}}$ using constant multiplicative factors (namely 0.391, 0.288, and 0.120, respectively) derived from the \citet{Cardelli_etal89} extinction curve for $R_V=3.1$. This simultaneous analysis would have been perfectly fine if the photometry were well calibrated, both from the data and stellar model sides. However, over the years we accumulated indications for the presence of small offsets in the VISTA photometry -- especially in the $Y$ band where the calibration is more problematic, owing to the absence of a $Y$-band in the calibrating data from 2MASS. 
One of these indications came from the detailed analysis of the best-fitting CMDs in \citet{Rub15}, for which the solutions appeared to be systematically shifted to the red in $\yks$ vs.\ $\ks$ diagrams, and to the blue in $\jks$ vs. $\ks$, by a few hundredths of a magnitude. These shifts may also have affected the extinction values derived by \citet{Rub15}, although a comparison with (widely varying) values in the literature did not reveal any evident problem -- apart from a likely overestimation of the \av\ values in the SMC outskirts, as compared to the \citet{Schlegel_etal98} maps. Early problems in the calibration of the $Y$ band were also reported by \citet{Rubele_etal12} and \citet{tatton13}\footnote{A new calibration of VISTA data \citep{gonzalez17} became available after we performed most of the analysis for this work. Its potential impact on our results is discussed in Appendix~\ref{app-calib}.}.
 
In view of this problem, we decided to decouple both solutions, searching for the best-fitting linear combination of partial models {\em independently} in the two CMD/Hess-diagrams. This means that two different solutions are found, characterized by slightly different SFHs, $\dmo$ and $\av$ values. We refer to these two solutions by the subscripts $Y\ks$ and $J\ks$, that is, 
     $[(m\mathrm{-}M)_0^{YK\mathrm{s}}, A_{V}^{YK\mathrm{s}}]$ 
and
     $[(m\mathrm{-}M)_0^{JK\mathrm{s}}, A_{V}^{JK\mathrm{s}}]$.
The best-fitting solution is also characterised by two $\chisqmin$ values, one for each CMD, which measure the residuals between the best-fitting model and the data. These values are presented in Table~\ref{tabellone}. Of course we can still define a total $\chisqmin$, as the sum of those derived from the two CMDs, i.e. $\chisqmin=\chisqmin_{,JK_\mathrm{s}}+\chisqmin_{,YK_\mathrm{s}}$. This latter value is used only to compare the present solutions with those obtained with the old method, where the two CMDs were analysed simultaneously. 

Since the present approach includes a new degree of freedom, it also decreases the total $\chisqmin$, improving the quality of the fitting. As shown in the example of Fig.~\ref{fig_sol}, for subregion G8 of tile SMC 5\_4, both CMDs are quite well fitted with this procedure, with residuals concentrated around the RC region of the CMD, but without any indication of a systematic colour mismatch between the data and the best-fitting model. There is also quite a good agreement between the SFH solutions derived from the two CMDs, as shown in the bottom right-hand panel of Fig.~\ref{fig_sfr}. On the other hand, the two best-fitting extinction values obtained for this region differ by $\avyk-\avjk \sim 0.133\pm0.067$~mag, which could be translated into a zero-point offset of about 0.05~mag, if interpreted as an offset in the $Y$ band only.
However, as can be verified from the numbers in Table~\ref{tabellone}, the $\avyk-\avjk$ difference varies significantly from tile to tile, ranging from $+0.31$ to $-0.17$~mag and with an average value of $0.06$~mag for the presently analysed area. Therefore, this problem cannot be simply attributed to a constant offset in the calibration, nor to a systematic error in the synthetic photometry \citep[see][]{Girardi_etal02} performed to build the stellar models.

Apart from the general improvement in the fitting this procedure also allows us to re-interpret the results using different extinction coefficients, since the $A_Y$, $A_J$ and $A_{K_\mathrm{s}}$ values -- now derived independently of each other -- could have been easily converted to $A_V$ using extinction curves different from the \citet{Cardelli_etal89} $R_V=3.1$ law. However, near-infrared extinction coefficients are little affected by changes in the interstellar extinction curve. We verified that this is the case for the range of extinction curves that can be expressed by means of \citet{gordon16}'s $A_\lambda/A_V[R_V,f_A]$ formalism: indeed, as $R_V$ is varied between 2 and 6 for $f_A=1$, and as $f_A$ is varied between 1 and 0 for $R_V=3.1$ -- that is, when the extinction curve is varied over the entire range observed inside the Milky Way, and from an average-Milky Way to the SMC one -- the maximum fractional variations in $A_Y/A_V$ are just 7 per cent. Even smaller are the variations in the $J$ and \ks\ bands. Therefore, the use of different extinction curves is unlikely to change the general interpretation of the data. 

The reader may also wonder why in Fig.~\ref{fig_sol}, the residuals concentrate around the RC in the CMD. The reason probably resides in the larger uncertainties of the evolutionary tracks at this stage of central helium burning, as compared to the main sequence and RGB phases. Indeed, the exact location and lifetime of RC stars is affected by uncertainties in the efficiency of core overshooting and its temperature gradient in the overshooting region, and by mass loss close to the tip of the RGB \citep[see][for a review]{girardi16}. Another feature that might be contributing to the larger residuals is the RGB bump, which for SMC metallicities appears very close to the RC, and which is sensitive to the assumed efficiency of envelope overshooting \citep[see][]{fu18}. Exploring these uncertainties is well beyond the scope of the present work.

It is also noteworthy that, in the subregion G11 of tile SMC 4\_2, the $\chisqmin$ turned out to be significantly larger than in neighbouring regions. This was due to the presence of 47~Tuc stars. Since the distance modulus of 47~Tuc is more than 5~mag shorter than the SMC, none of the available partial models could provide any significant match to the distribution of its stars in the CMD. Therefore the derived best-fitting results appear not having been affected at a significant level.

\subsection{Evaluating depth effects}
\label{sec:depth}

\begin{figure}
\centering
\resizebox{0.49\hsize}{!}{\includegraphics{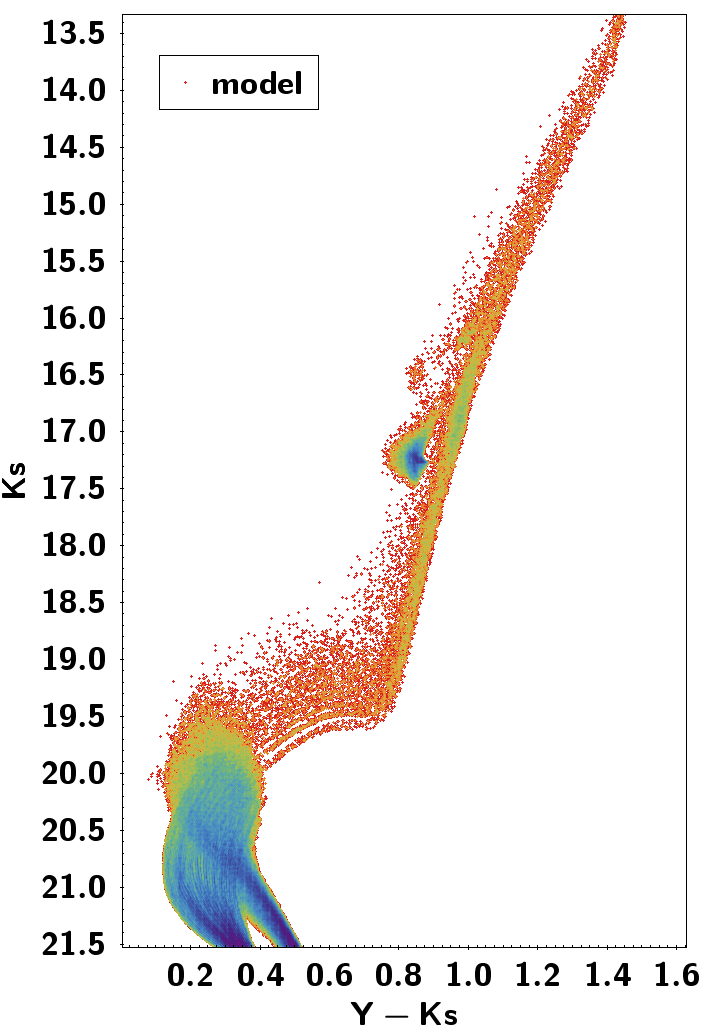}}
\resizebox{0.49\hsize}{!}{\includegraphics{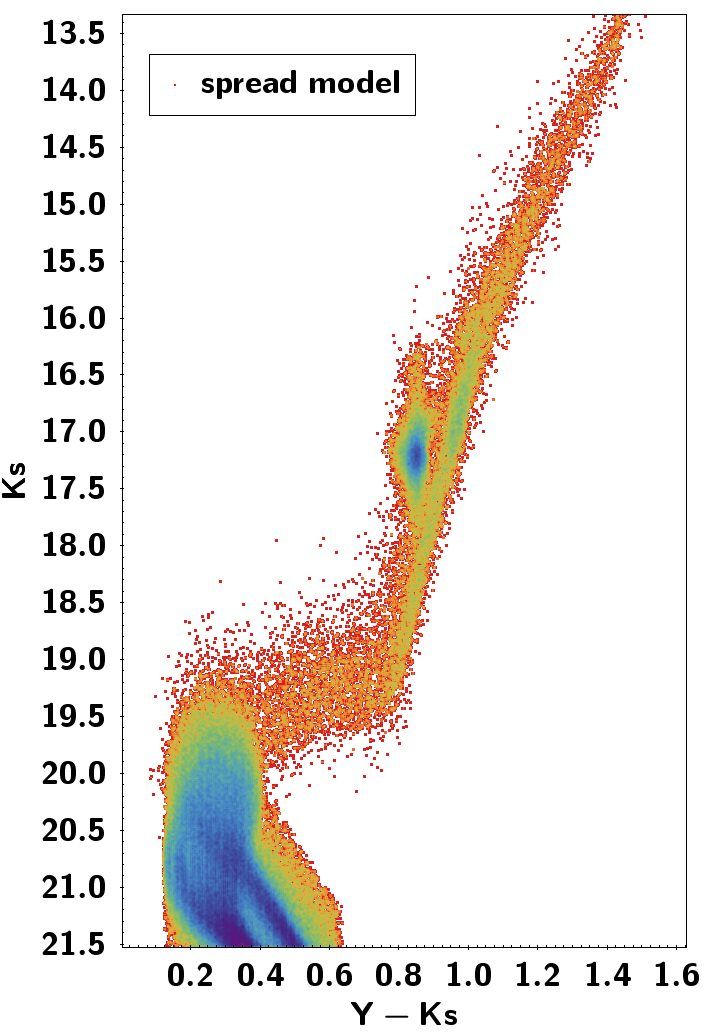}}
\\
\resizebox{0.7\hsize}{!}{\includegraphics{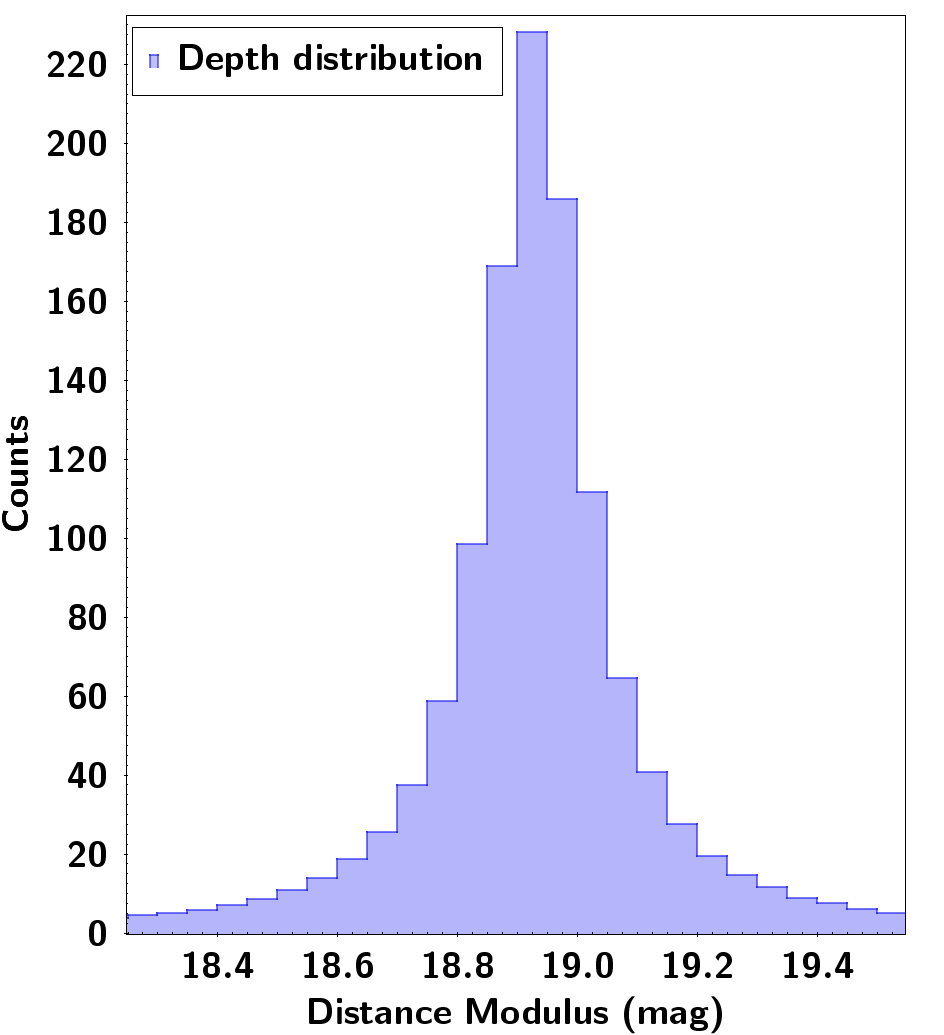}}
\\
\resizebox{0.9\hsize}{!}{\includegraphics{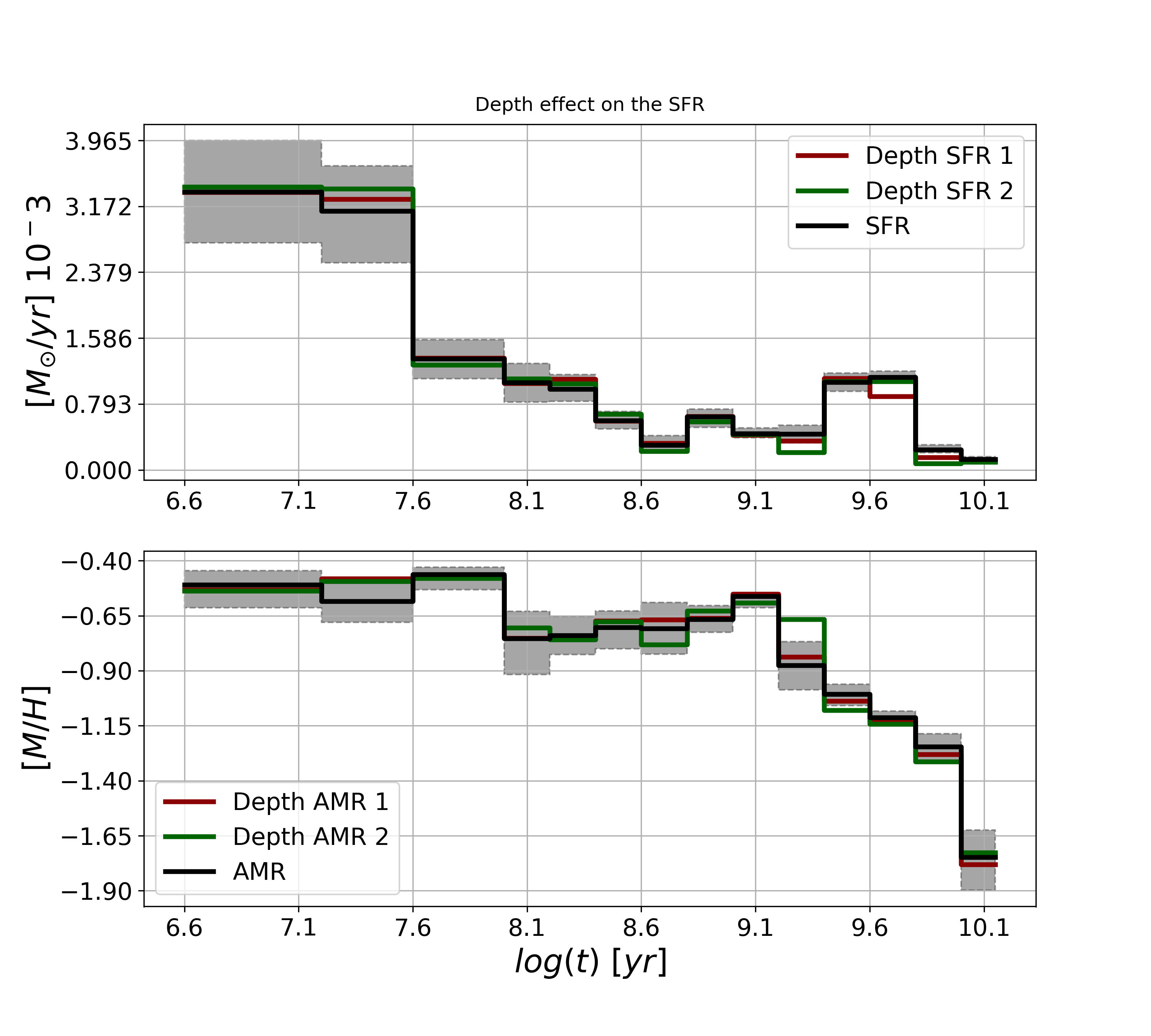}}
\caption{Example of the SMC distance depth effect applied to a CMD (\ks\ versus \yks) of a partial model with $\log(t/\mathrm{yr})=9.1$, and their effect on the SFR reconstruction.  Top panels show the simulated CMD for the zero-depth case, and its depth-corrected version. The middle panel shows the assumed depth distribution. The bottom panels show the effect on the derived SFR and AMR.
The black line and gray-shaded area correspond to the zero-depth best-fitting solution (the same as in Fig.~\ref{fig_sfr}), while the green and red lines are solutions that include the depth model: The green line is the solution at the same distance and extinction as the zero-depth case, while the red one is the solution for the new best-fitting value of distance and extinction.}
\label{fig.depth}
\end{figure}

As already noted, SMC populations show clear indications of structures along the line of sight, whose depths vary considerably depending on the SMC region, and on the distance tracer used. \citet{HZ04} already assessed the impact that a significant depth would have on the derivation of the SFH, by producing a 12~kpc deep synthetic model of the SMC and analysing it without taking into account the spread in distance. They concluded that the recovered SFHs were the same as the input ones, within the errors. 

Here, we perform a similar test, but the other way around as compared to the \citet{HZ04} one: we analyse the same VMC data adopting synthetic populations located either at a single distance (zero depth) as before, or spread according to a depth distribution. The first step in this exercise is to produce partial models that include a depth compatible with the observed data. This is illustrated in the top panels of Fig.~\ref{fig.depth}, which show the same partial model with a zero depth (left panel), and after assuming a depth distribution (central panel) which resembles very much the one derived by \citet{Muraveva18} for the RR~Lyrae stars in the SMC. The latter is shown in the top right-hand panel of the figure; it was created by a Cauchy function with $\gamma=2.5$~kpc. The full width at half maximum of this distribution slightly exceeds the average depth of $4.3\pm1.0$~kpc derived by \citet{Muraveva18}, and yet it includes extended tails in the distance distribution, reaching total depths as large as 25~kpc.

Using similarly derived partial models for all ages and metallicities, we perform the SFH recovery exactly in the same way as before. The results for subregion G8 of tile SMC 5\_4, in terms of SFR$(t)$ and $\mh(t)$, are shown in the bottom panels of Fig.~\ref{fig.depth}: The dark lines with gray-shaded areas show the solution obtained with the standard zero-depth method, and its error bars (as already shown in Fig.~\ref{fig_sfr}). The green line instead is the best-fitting solution found after assuming a depth, for the same values of \av\ and \dmo\ as in the zero-depth method. Finally, the red line is a slightly better solution, found after exploring the depth solutions over a grid of \av\ and \dmo\ values, so as to redetermine their best-fitting values. As can be seen, the differences among these three solutions are almost imperceptible, and usually within the error bars of the zero-depth solution. Therefore, we reach the same conclusion as \citet{HZ04}, that the zero-depth solutions are essentially the same as those found assuming a reasonable depth distribution. Of course this aspect of the method must be improved once we have more definitive indications about the distance distributions to be adopted for populations of different ages, in different parts of the SMC \citep[as those from][Tatton et al.\ in prep., for Cepheids, RR Lyrae and RC stars in VMC, respectively]{ripepi17,Muraveva18}.

\section{The extinction and distance distribution in the SMC}
\label{sec:D_AV}
\begin{figure*}
\resizebox{0.33\hsize}{!}{\includegraphics{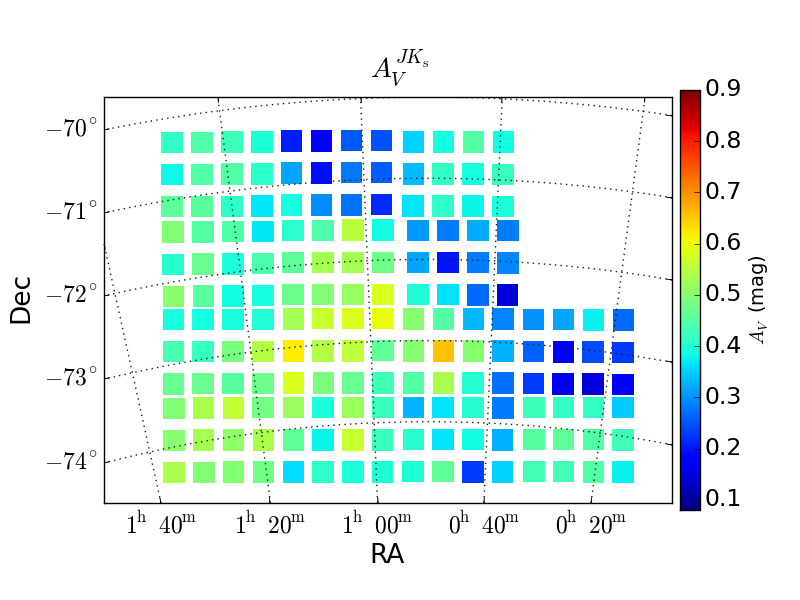}}
\resizebox{0.33\hsize}{!}{\includegraphics{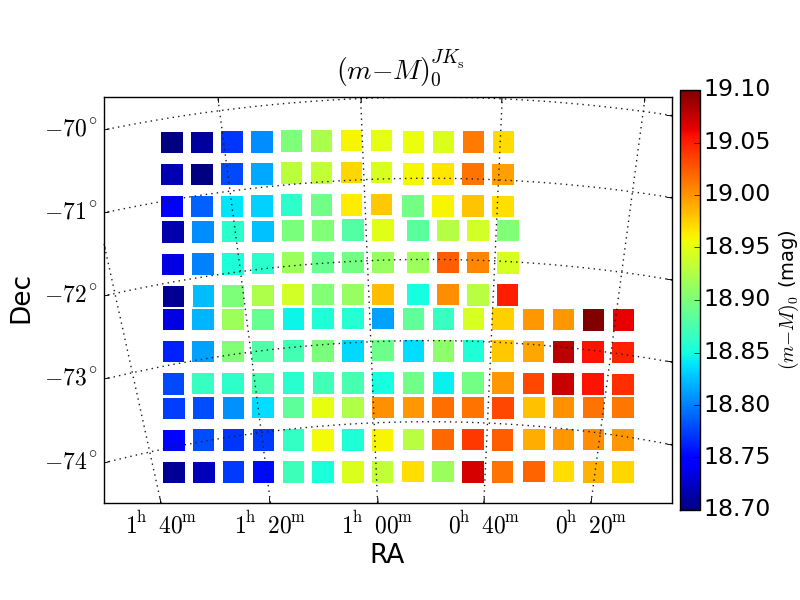}} \\
\resizebox{0.33\hsize}{!}{\includegraphics{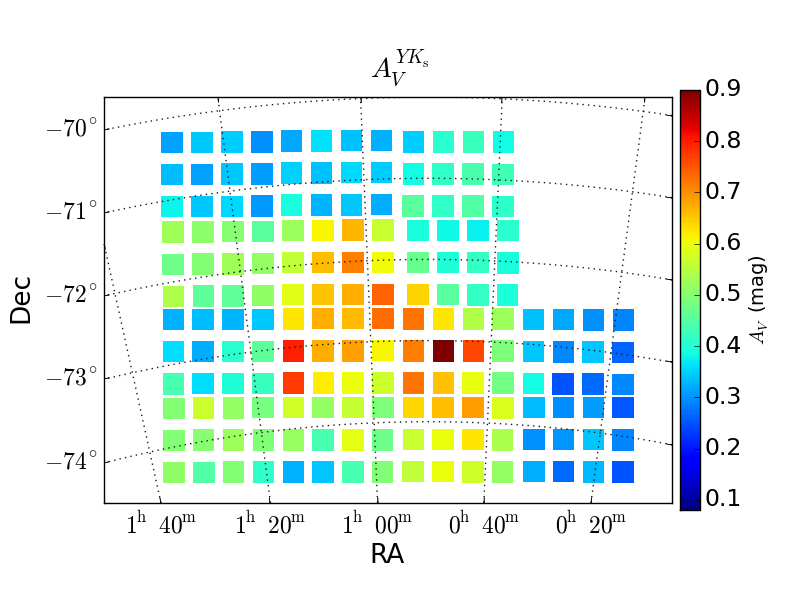}}
\resizebox{0.33\hsize}{!}{\includegraphics{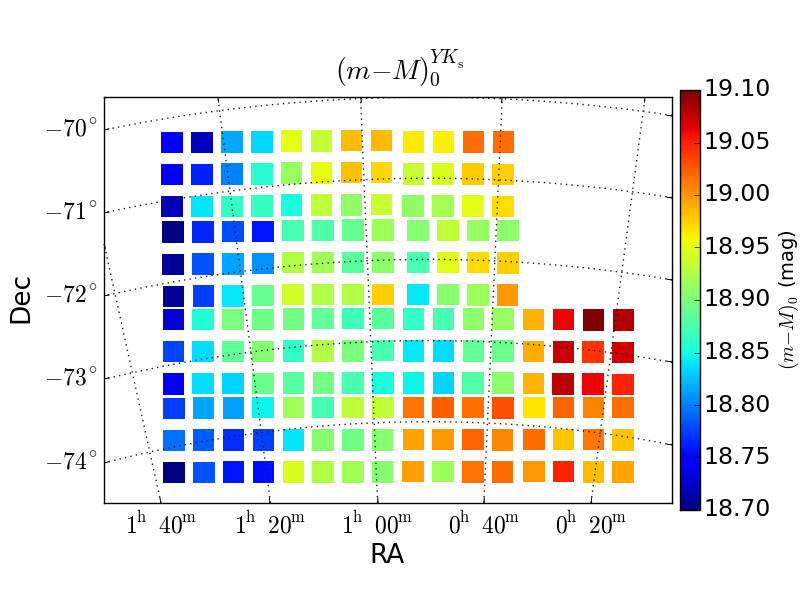}}
\caption{Spatial distributions of the extinction (left panels) and true distance modulus (right-hand panels), derived either using the \jks\ vs.\ \ks\ CMD (top panels), or \yks\ vs.\ \ks\ (bottom panels).}
\label{fig.AVD}
\end{figure*}

As described by \citet{Rub15} and recalled in Sect.~\ref{sec:method}, in our pipeline the parameters \av\ and \dmo\ are considered free variables in the minimisation process. Therefore for each of the 168 subregions analysed we recover the best-fitting extinction and distance as additional outputs of the SFR and AMR. In addition, we have two values for these parameters, one for each CMD. The best-fitting values are identified by fitting a second-order polynomial to the $\chisq$ distribution across the $\av$ versus $\dmo$ plane, as illustrated in Fig.~\ref{fig_sfr}. Error bars are derived from synthetic realisations of the best-fitting models, as explained by \citet{Rubele_etal12}. Fig.~\ref{fig.AVD} shows the spatial distribution of these values as a function of right ascension (RA) and declination (Dec). Panels at the top and bottom rows present the values obtained using \jks\ vs.\ \ks\ and \yks\ vs.\ \ks\ CMDs, respectively.

\subsection{Extinction: results and comparisons}
As one can appreciate from the extinction maps of Fig.~\ref{fig.AVD}, the derived \av\ varies between 0.1 and $\sim0.9$~mag across the SMC, with the smallest values in the external regions and the highest values concentrated in a triangle-shaped region that coincides with the SMC Bar plus Wing. It is also clear that the high extinction values follow the distribution of the youngest stellar populations (see the panels with ages $\log(t/\mathrm{yr})<9$ in Figs.~\ref{SFRmap} and \ref{MASSmap}), as well as the H\textsc{i} distribution shown by \cite{Stani2004}.

Since the \jks\ vs.\ \ks\ CMDs do not involve any significant adjustment in the photometry
we consider that they provide the most reliable \av\ values. The average extinction we found based on all SMC regions analysed is $\avjk = 0.41$~mag ($0.46$~mag is the average value of $\avyk$). The outskirts towards the North and South-West of the SMC present $\avjk<0.3$~mag, while in the centre of the galaxy and along the direction of the Bridge the typical values are $\sim\!0.5$~mag, with maximum values of $\sim\!0.7$~mag.

While these results are globally consistent with our previous work, we note that the $\avjk$ values derived in this work are systematically smaller than the $\av$ values of \citet{Rub15} by about 0.06~mag. 
That said, there are a few aspects which are clearer among the present results: 
\begin{itemize}
\item The external regions of the SMC present values of $\avjk\sim0.25$~mag, which are larger than the $\av \sim 0.12 $~mag derived from the \cite{Schlegel_etal98} maps. 
\item Similar extinction maps were derived by \cite{Israel2010} using WMAP and COBE data. They provide a mean extinction of about $\av=0.45$~mag internal to the SMC; these became $\av\sim 0.6$~mag when the Milky Way extinction is added. The results by \cite{Israel2010} seem to exceed our extinctions by more than $\sim\!0.2$~mag. 
\item Conversely our average \avjk\ is in good agreement with the values derived by \citet{Zaritsky_etal2002}, which range between 0.15~mag and 0.65~mag when we consider respectively cool and hot stars. This work is the only large-area survey of the SMC that has estimated the extinction star-by-star (see also Tatton et al., in preparation).
\item There is a significant diference between our extinction values and those obtained by \citet{Haschke11} and \citet{Subramanian2012}. The former work uses the OGLE III survey database to obtain an average $\av \simeq 0.1 \pm 0.15$ mag from the RC stars de-reddening method, and  $\av \simeq 0.18 \pm 0.15$ mag from RR Lyrae stars. Similar results were found by \citet{Subramanian2012} using the same methods, and by \citet{Muraveva18} using RR Lyrae stars in the SMC observed by the VMC survey. 
\end{itemize}  

\subsection{Distance: results and comparisons}

Since the values derived for \dmo\ from the $J\ks$ and $Y\ks$ CMDs generally agree within the errors, in the following we will refer to the mean values.

For the coordinates of the SMC centre derived in our previous work (see figure 9 in \citealt{Rub15}) at $\alpha=12.60^\circ$, $\delta=-73.09^\circ$, we derive the new distance for the centre of the SMC $\dmo=18.863 \pm 0.023$~mag ($d=59.24$~kpc), just about 0.2~kpc farther than the value found by \citet{Rub15}. 
This value is smaller than the values favoured based on Cepheids from the VMC survey \citep[either $19.01\pm0.05$ or $19.04\pm0.06$;][]{ripepi15}. On the other hand, our distances are within the wide range of values obtained by many different methods in the literature (see \citealt{degrijs15}, for a review), and especially those based on the RC mean magnitude (namely $\dmo=18.88\pm0.03$ with a standard deviation of 0.08~mag; \citealt{degrijs15}). They also agree with the weighted mean distance modulus of $\dmo=18.88\pm0.20$~mag found by \citet{Muraveva18} for the SMC old stellar component as traced by 2997 RR~Lyrae variables observed by the VMC survey. We recall that our distances follow from the direct comparison between the photometry of stellar models and the data, and lack an independent calibration based on primary standard candles. Therefore, although the agreement with distance estimates based on intermediate-age and old tracers like RC stars and RR~Lyrae is encouraging, our mean value of the true distance modulus can still be affected by (hard to assess) systematic errors.

The right-hand panels in Fig.~\ref{fig.AVD} show maps of the true distance modulus as a function of the coordinates, as derived from the two CMDs. It is evident that the Eastearn and South-eastern regions, in the direction of the Magellanic Bridge, correspond to the closest part of the SMC galaxy, whereas the South-western part is the farthest. As discussed by \citet{Rub15}, these trends are in agreement with many recent works. For instance, using the RC method \cite{Subramanian2012} find similar trends as regards the differential distance. Also \citet{deb17}, \citet{jacy17} and  \citet{Muraveva18}, using RR~Lyrae stars, confirm the trend of increasing distances as one goes from the Southeast to the Southwest of the SMC. On the other hand, Cepheid periods provide a somewhat different picture for the young (50-500 Myr) populations: they are found to have a significant 3D structure and depths exceeding 20~kpc \citep[see][]{ripepi17}. 

We also re-determine the centre of mass of the galaxy as the weighted mean of the coordinates and distances of all subregions, using the mass assembled in each sub-region (see Fig.~\ref{MASSmap}) as the weights. With this method we find the centre at $\alpha=13.32^\circ \pm 1.10$, 
$\delta=-72.93^\circ \pm 0.86$ and $ \dmo=18.910 \pm 0.064 $~mag. These coordinates agree with the centre derived using star counts by \citet{Rub15}, and they are in the region in which the stellar density varies by less than $10$ per cent. 

\section{The star formation history and chemical evolution}
\label{sec:SFRAMR}

Similarly to \citet{Rub15}, we define the total SFR$(t)$ as the sum of the SFR values derived for each age bin, and the mean metallicity $\mh(t)$ (or age--metallicity relation, AMR) as the weighted mean of \mh\ for each age bin, with the weight provided by the SFR found for each partial model. Also the error bars are defined in a similar way, by properly weighing the errors found for each partial model. The fact that we have solutions from two different CMDs does not change the analysis significantly: as a rule, both solutions agree within their 1$\sigma$ error bars (see the example in Fig.~\ref{fig_sfr}). Therefore, in the following we simply adopt the mean of the two solutions.
 
\subsection{Maps of the SFR}
\label{sec:sfr}
 
\begin{figure*}
\vspace{-0.5cm}
\resizebox{0.33\hsize}{!}{\includegraphics{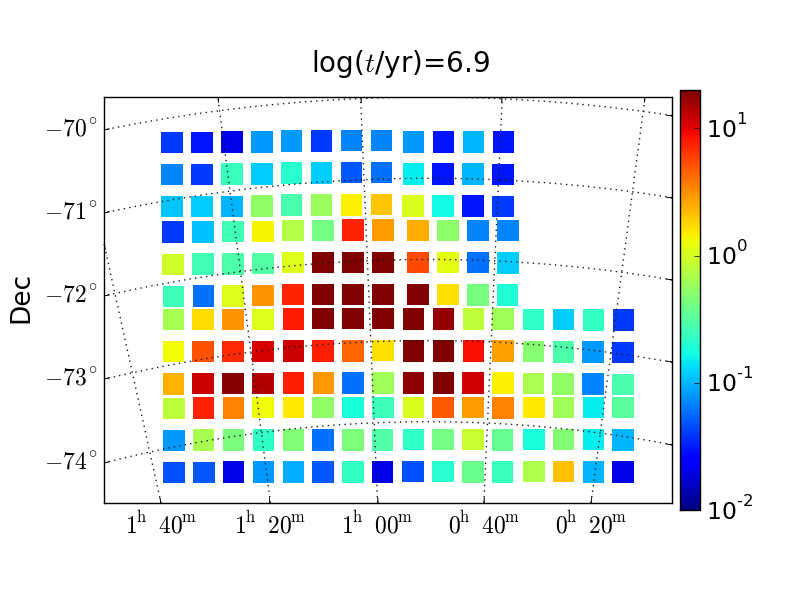}}
\resizebox{0.33\hsize}{!}{\includegraphics{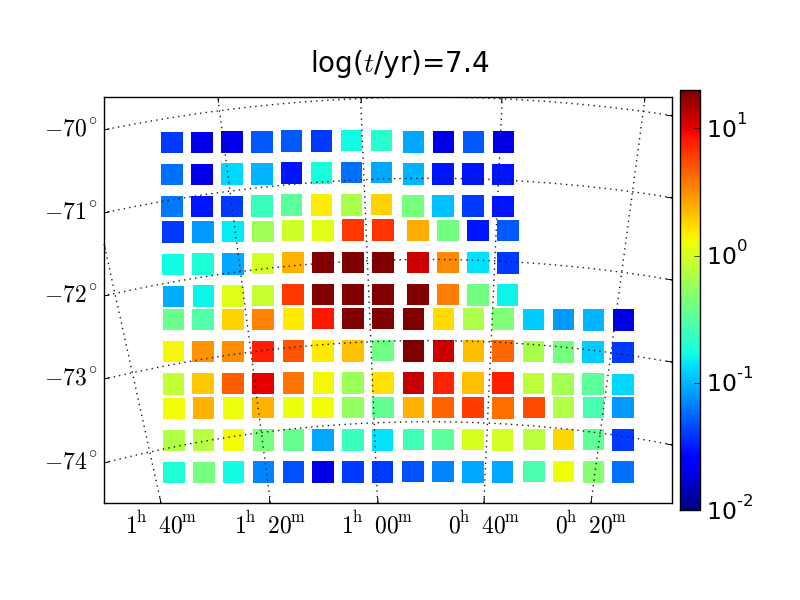}}
\resizebox{0.33\hsize}{!}{\includegraphics{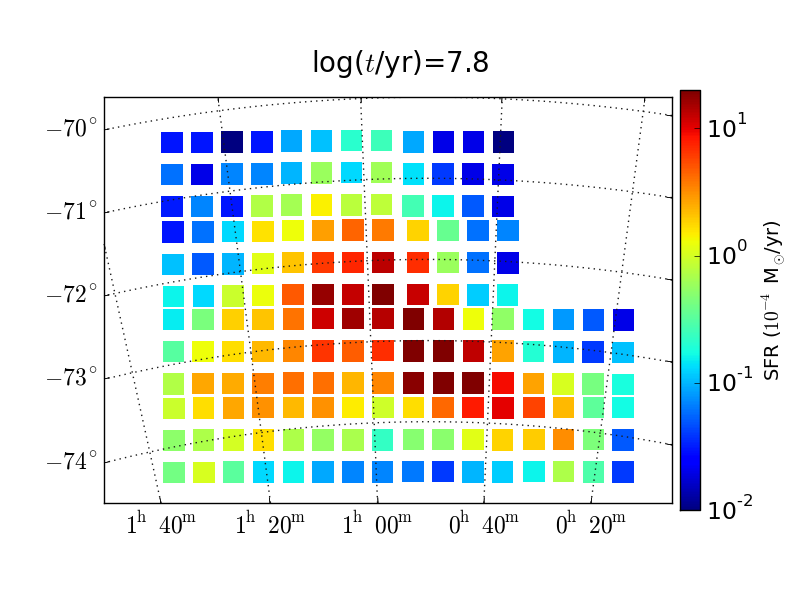}}\hfill\\
\resizebox{0.33\hsize}{!}{\includegraphics{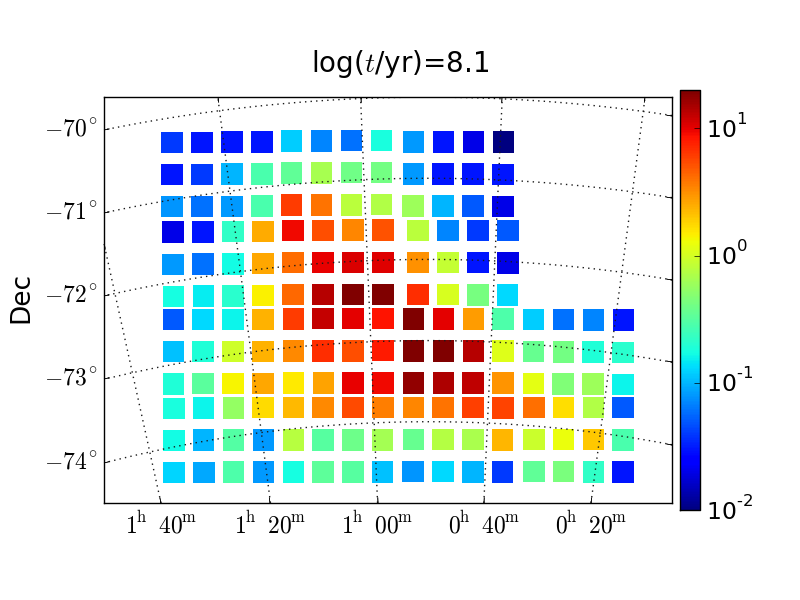}}
\resizebox{0.33\hsize}{!}{\includegraphics{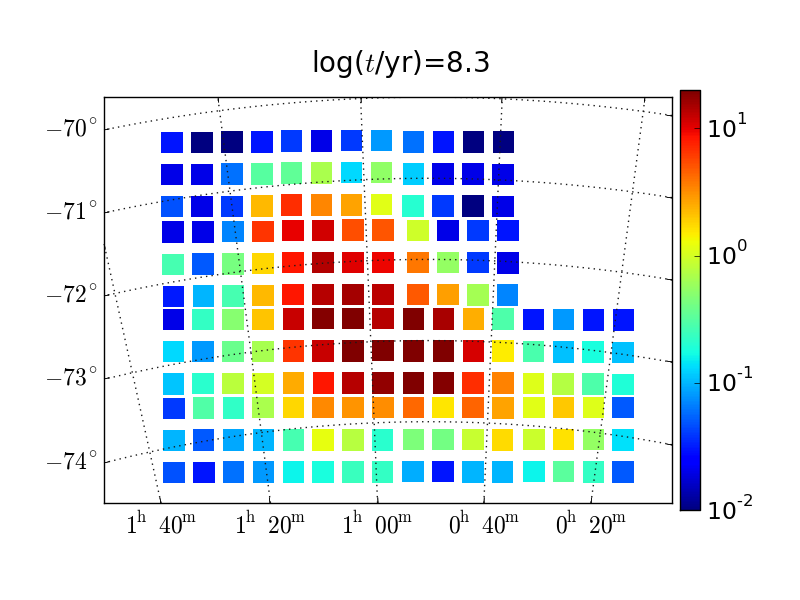}}
\resizebox{0.33\hsize}{!}{\includegraphics{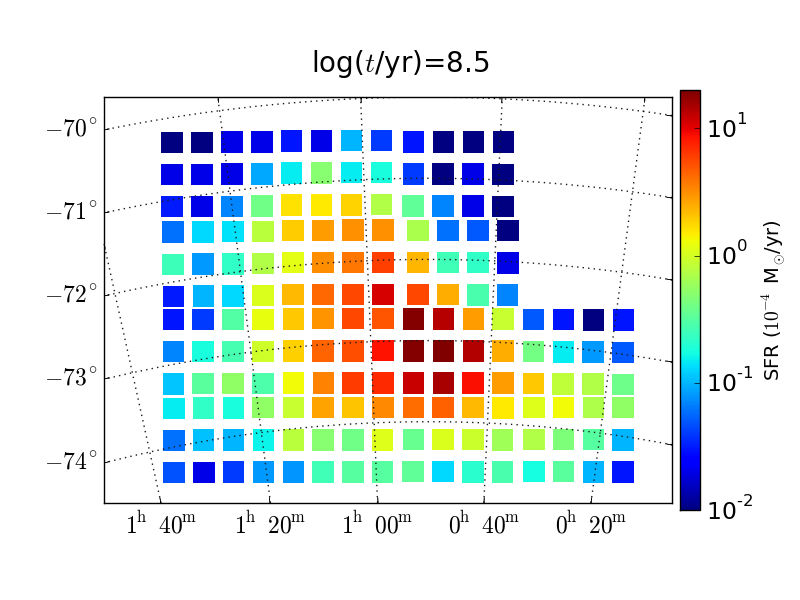}}\hfill\\
\resizebox{0.33\hsize}{!}{\includegraphics{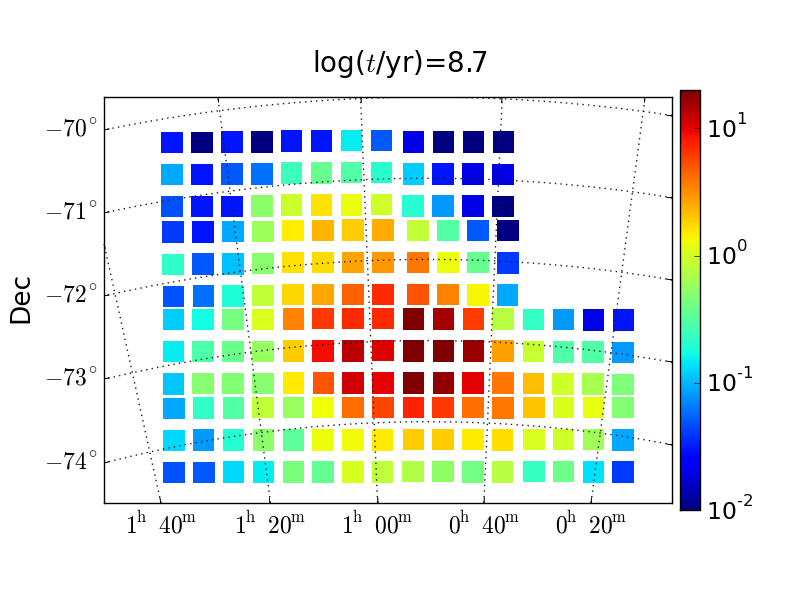}}
\resizebox{0.33\hsize}{!}{\includegraphics{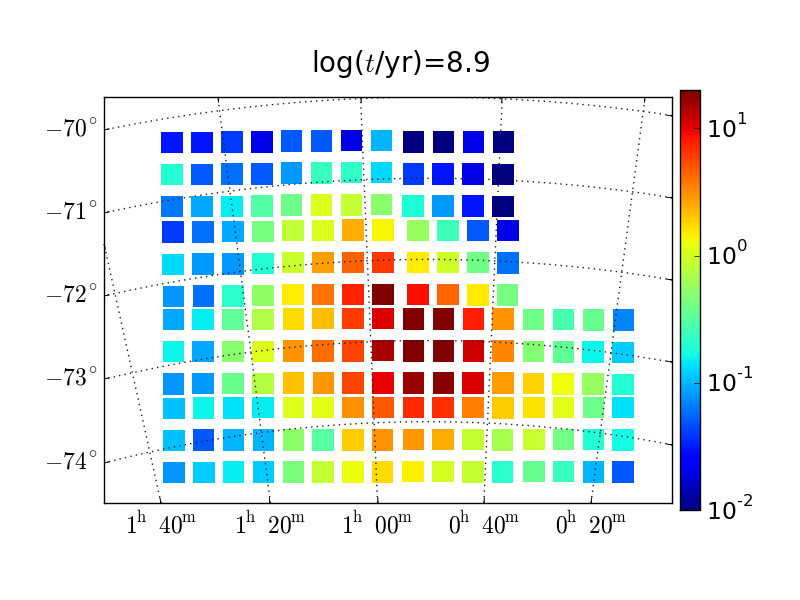}}
\resizebox{0.33\hsize}{!}{\includegraphics{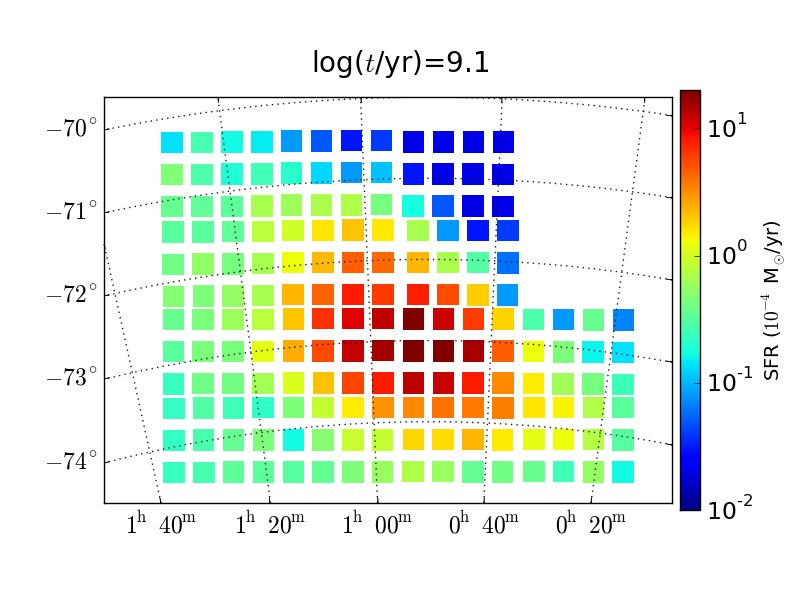}}\hfill\\
\resizebox{0.33\hsize}{!}{\includegraphics{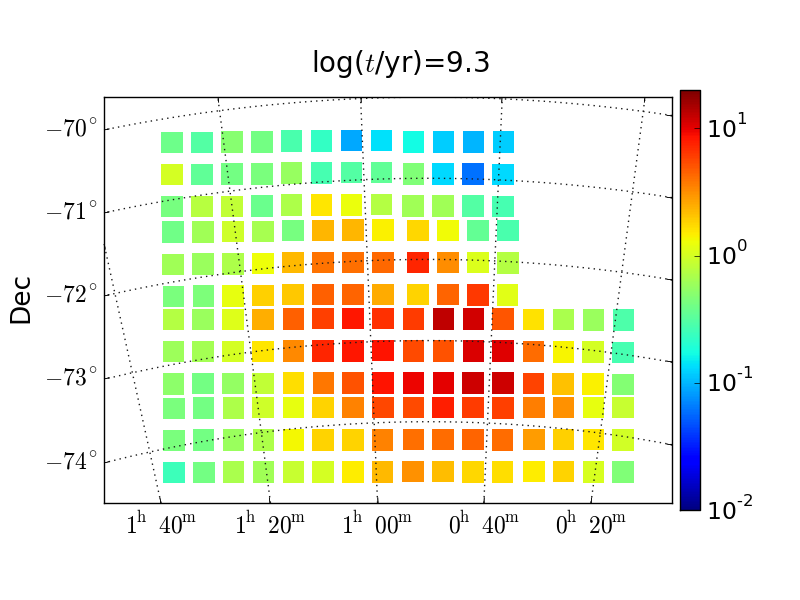}}
\resizebox{0.33\hsize}{!}{\includegraphics{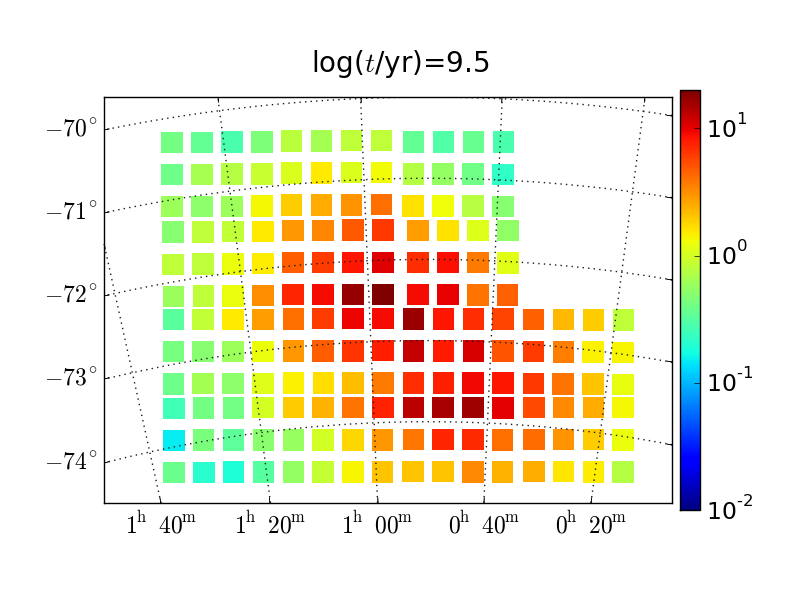}}
\resizebox{0.33\hsize}{!}{\includegraphics{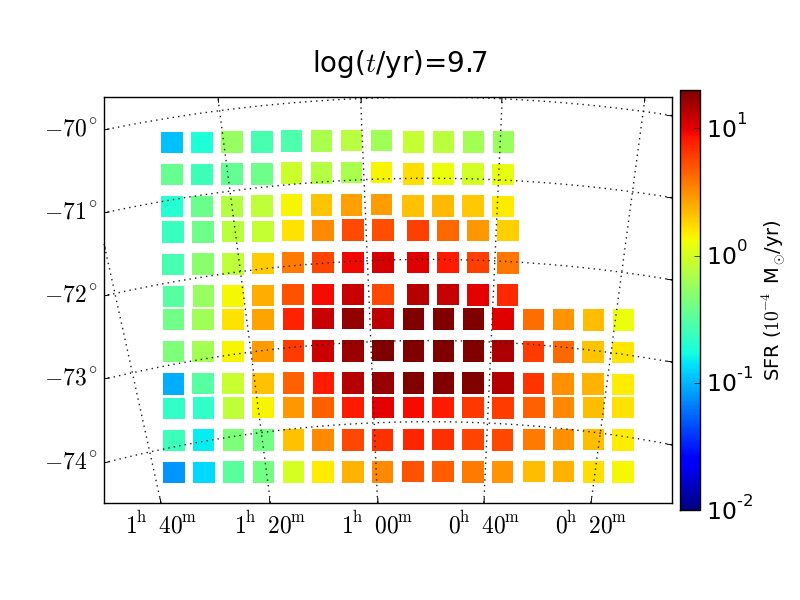}}\hfill\\
\resizebox{0.33\hsize}{!}{\includegraphics{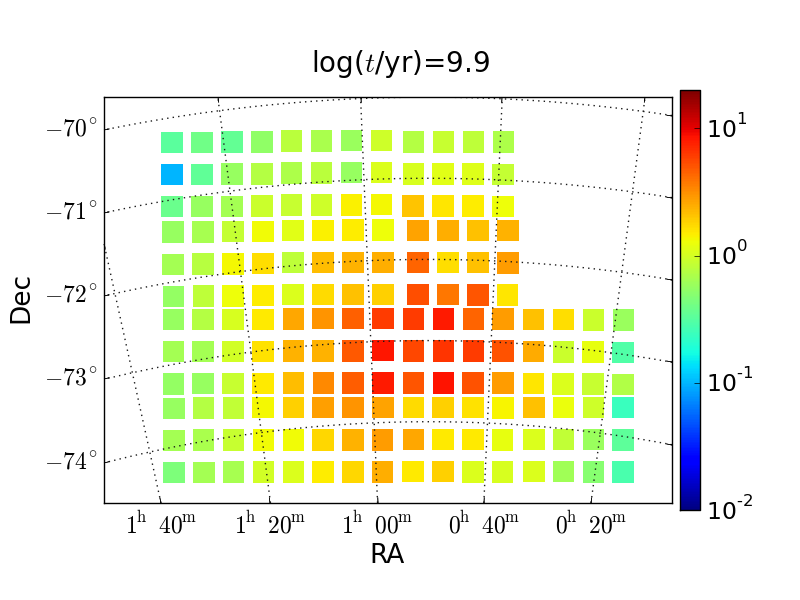}}
\resizebox{0.33\hsize}{!}{\includegraphics{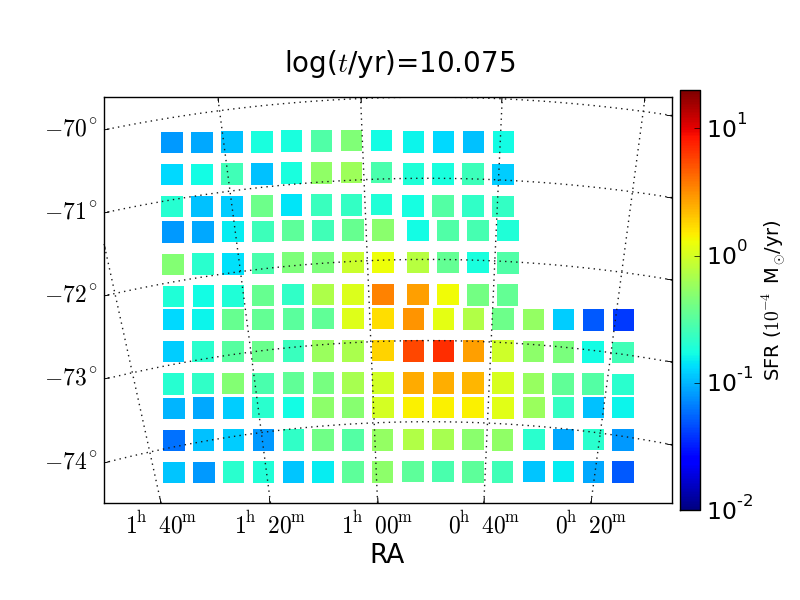}}
\resizebox{0.33\hsize}{!}{\includegraphics{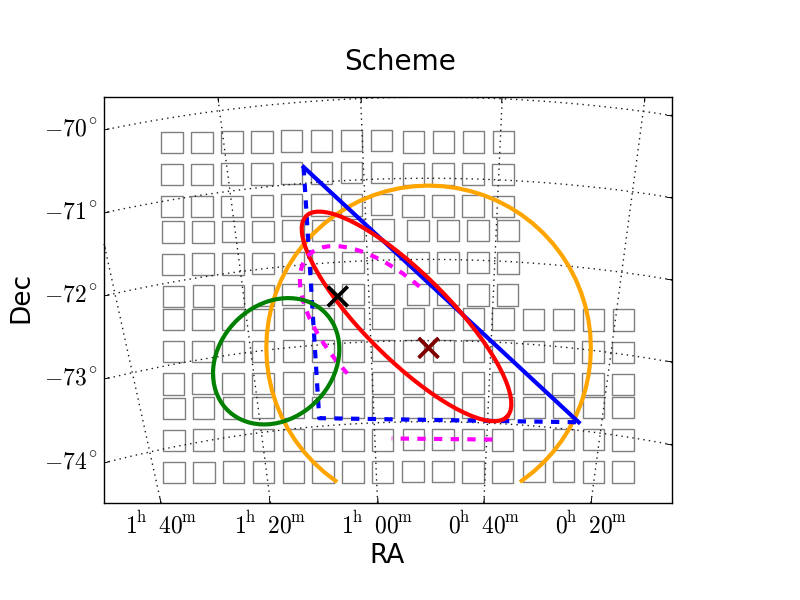}}\hfill
\caption{SFR across the SMC galaxy as a function of age, from the youngest (top-left panel) to the oldest (bottom) age bin considered in this work. Each square corresponds to a subregion, with colours indicating the SFR in units of $10^{-4}$~$\Msun\,\mathrm{yr}^{-1}$. The bottom right-hand panel indicates, very schematically, the position of the features mentioned in the text: the Bar (red ellipse) and the Wing (green ellipse) defined by young populations ($\log(t/\mathrm{yr})<7.8$), the triangular region that encompasses most of the $7.8<\log(t/\mathrm{yr})<8.5$ populations (blue lines) together to its marked Northwestern edge (blue continuous line), the kinematical centre by \citet[][black cross]{stani04}, the star counts centre \citet[][maroon cross]{Rub15} together with a circular area or radius $2\degr$ (orange circle) that encompasses most of the stars formed in the intermediate-age-to-old spheroid ($\log(t/\mathrm{yr})\gtrsim9.7$), and finally the ``on'' populations used to define the 2.5~Gyr ring by \citet[][magenta dashed lines]{HZ04}. 
}
\label{SFRmap}
\end{figure*}

\begin{figure*}
\resizebox{0.33\hsize}{!}{\includegraphics{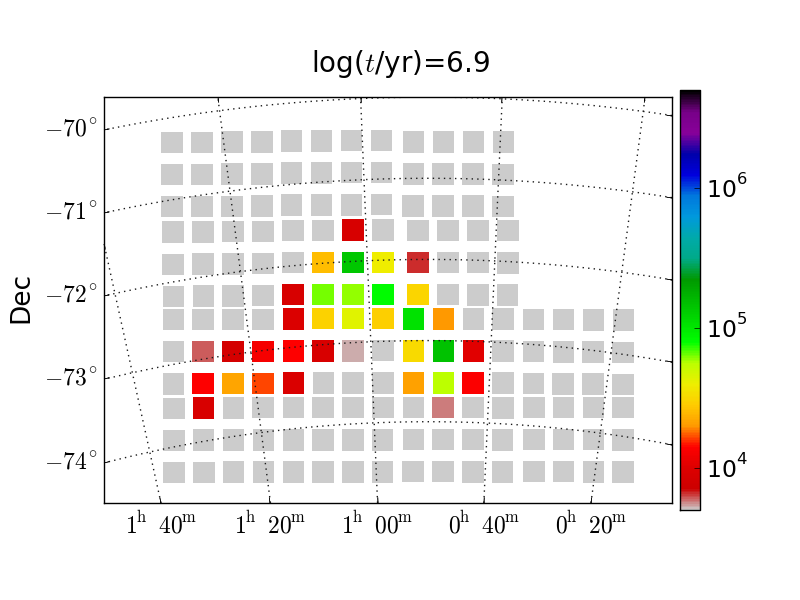}}
\resizebox{0.33\hsize}{!}{\includegraphics{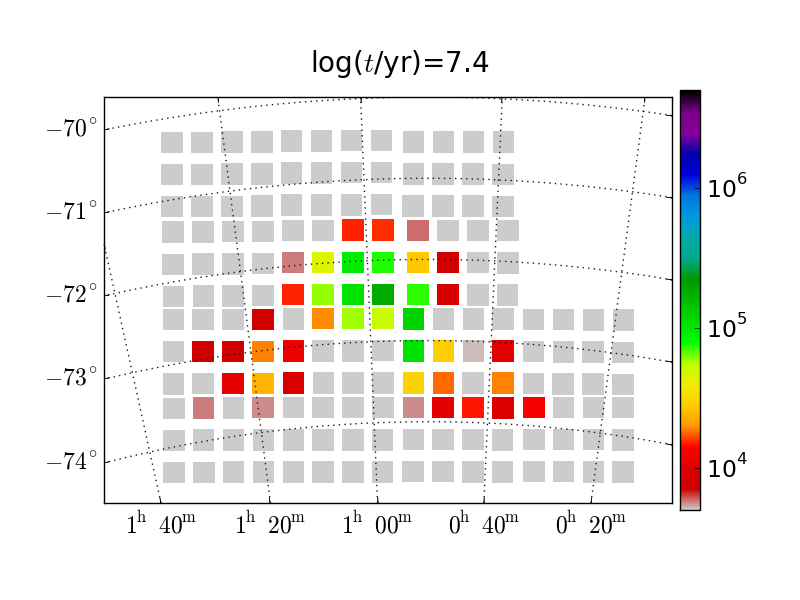}}
\resizebox{0.33\hsize}{!}{\includegraphics{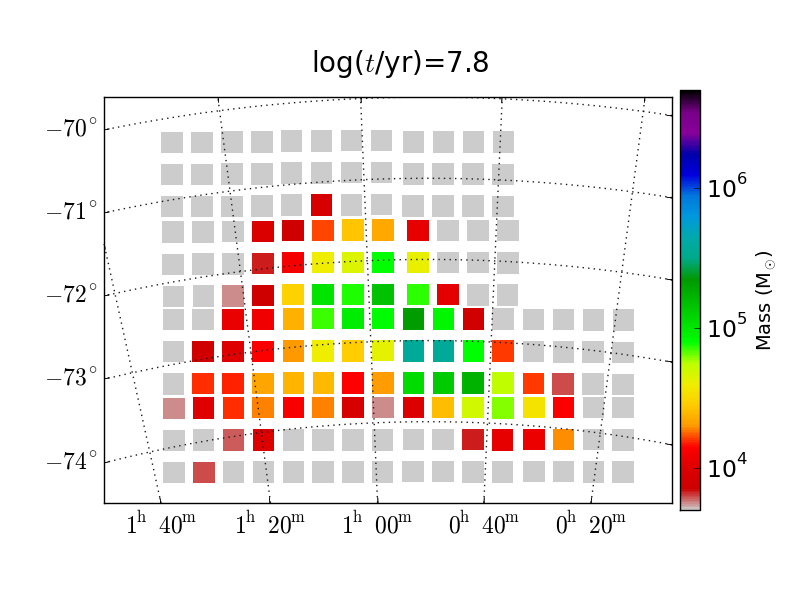}}\hfill\\
\resizebox{0.33\hsize}{!}{\includegraphics{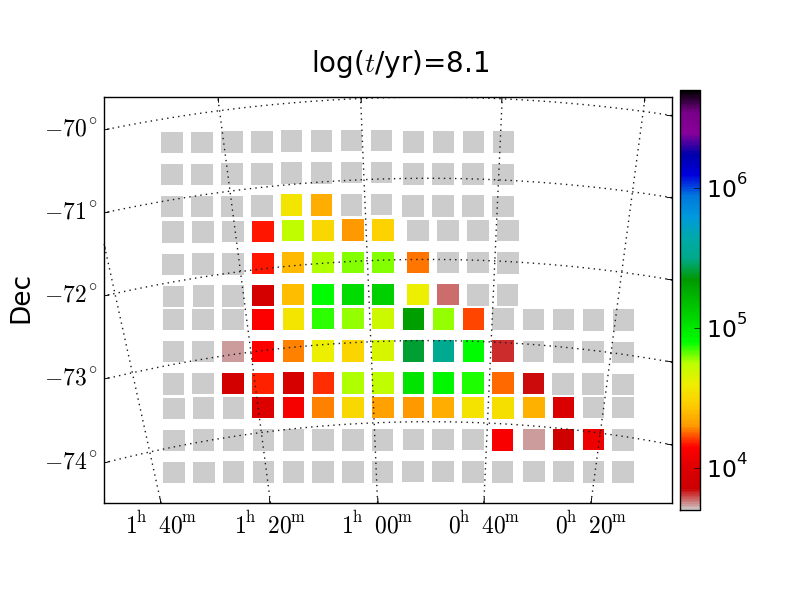}}
\resizebox{0.33\hsize}{!}{\includegraphics{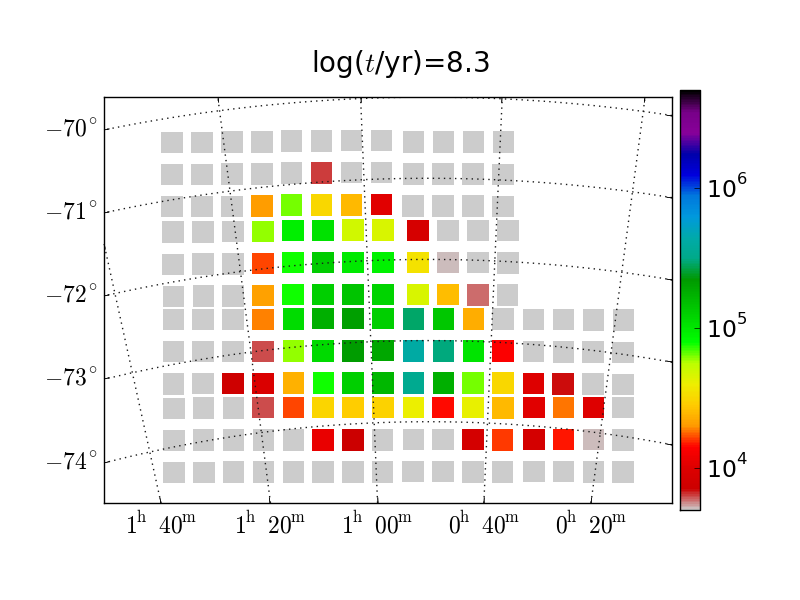}}
\resizebox{0.33\hsize}{!}{\includegraphics{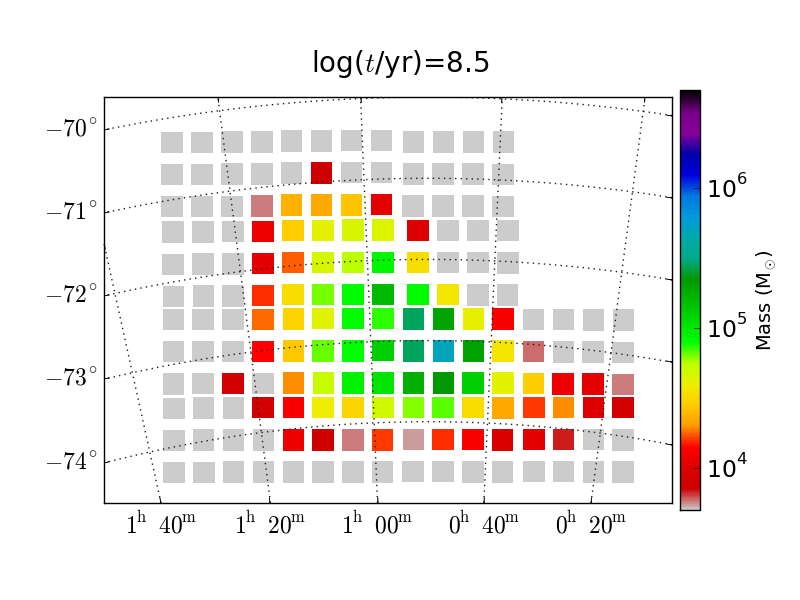}}\hfill\\
\resizebox{0.33\hsize}{!}{\includegraphics{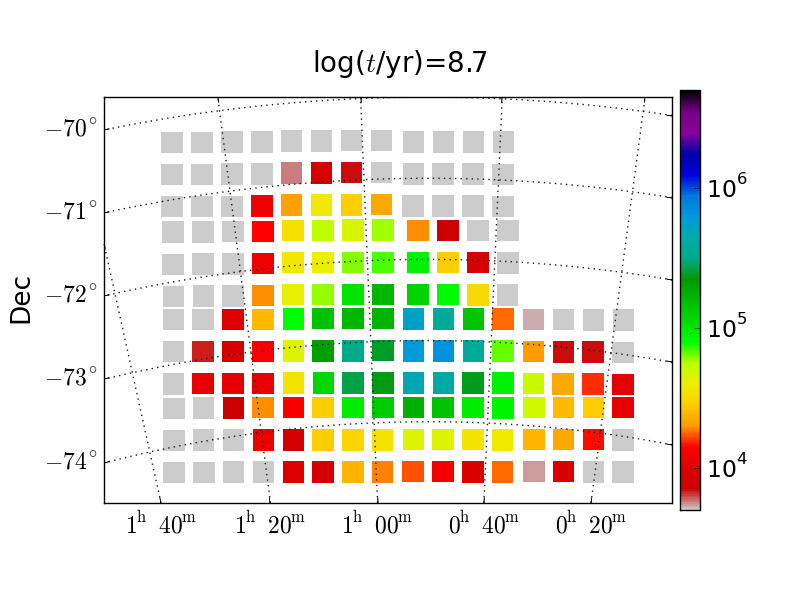}}
\resizebox{0.33\hsize}{!}{\includegraphics{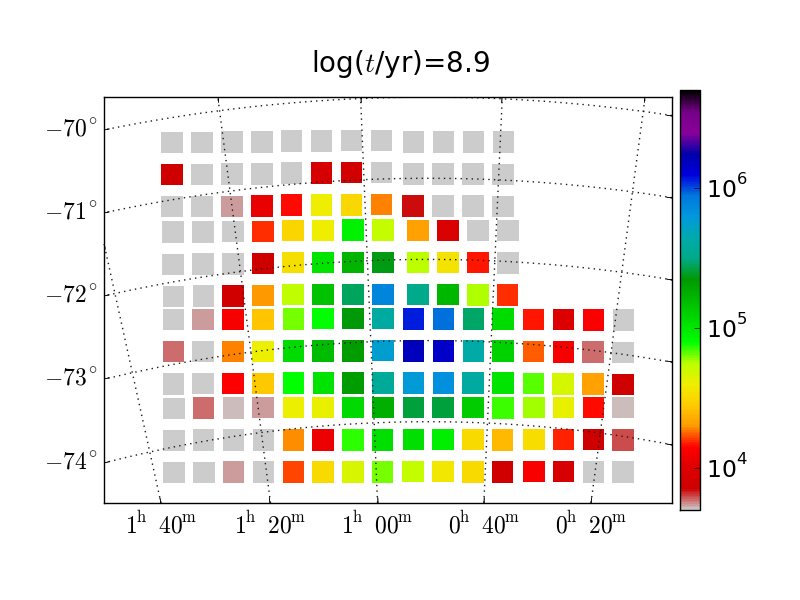}}
\resizebox{0.33\hsize}{!}{\includegraphics{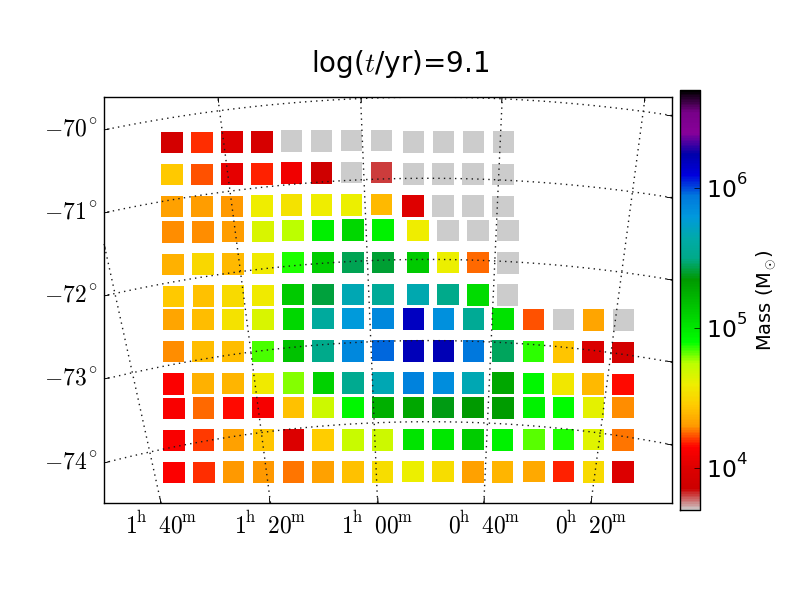}}\hfill\\
\resizebox{0.33\hsize}{!}{\includegraphics{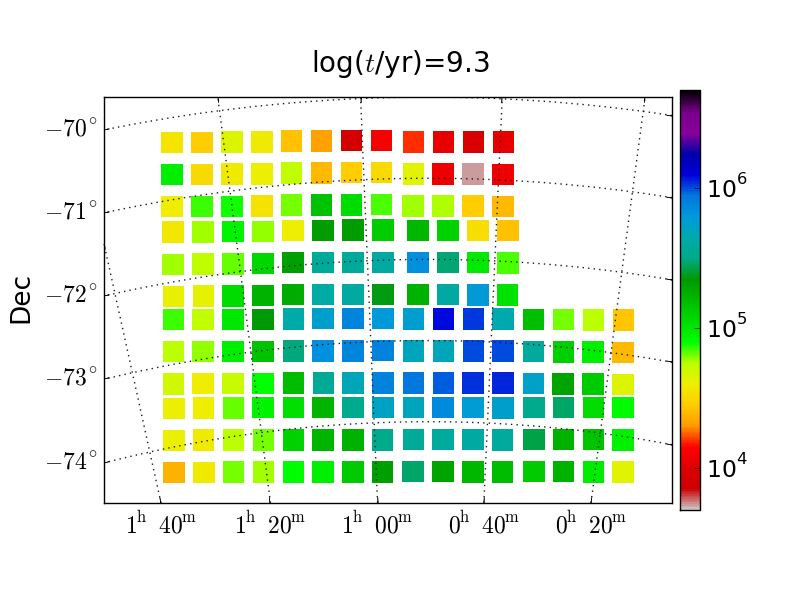}}
\resizebox{0.33\hsize}{!}{\includegraphics{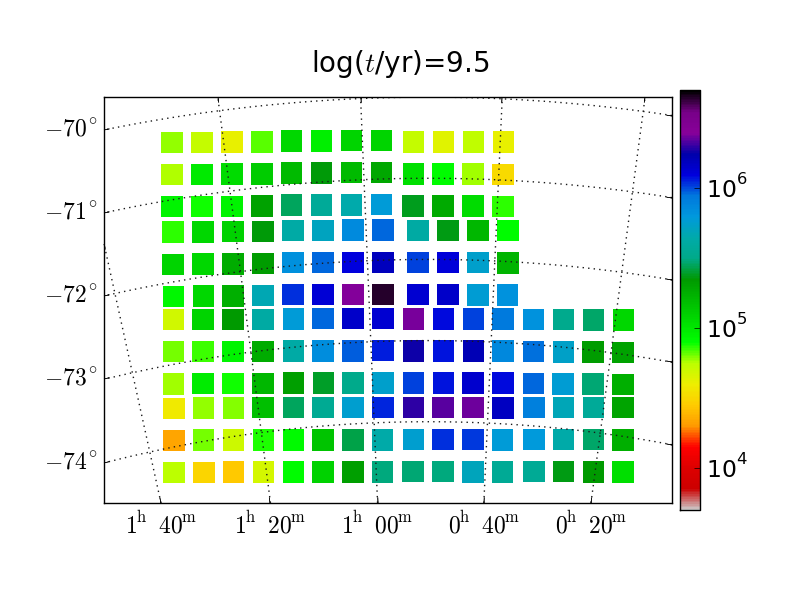}}
\resizebox{0.33\hsize}{!}{\includegraphics{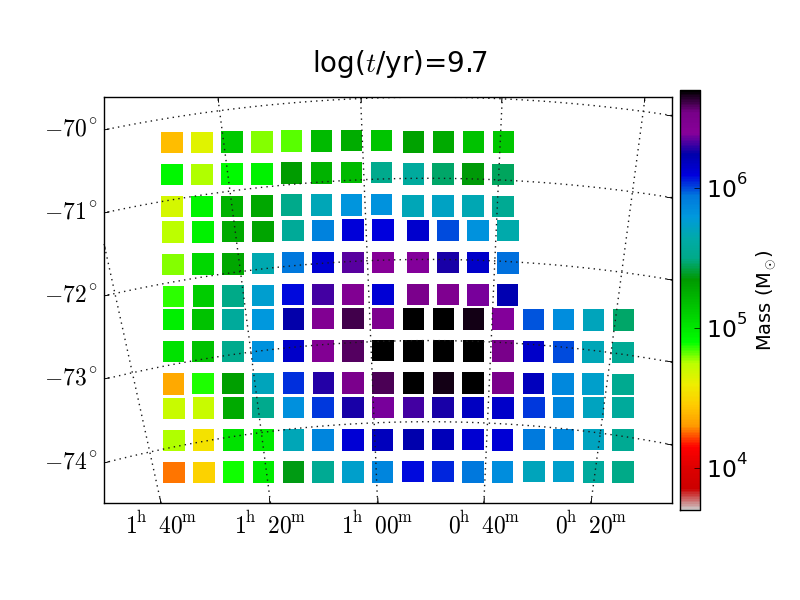}}\hfill\\
\resizebox{0.33\hsize}{!}{\includegraphics{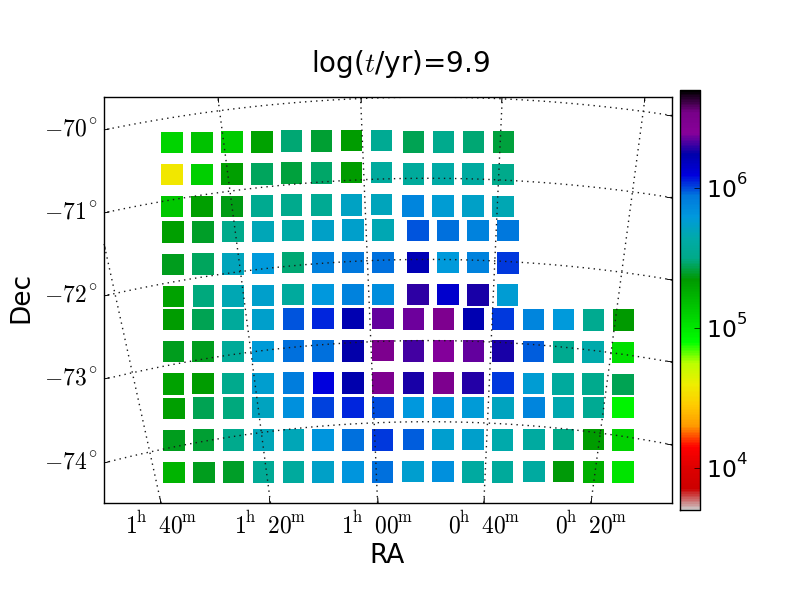}}
\resizebox{0.33\hsize}{!}{\includegraphics{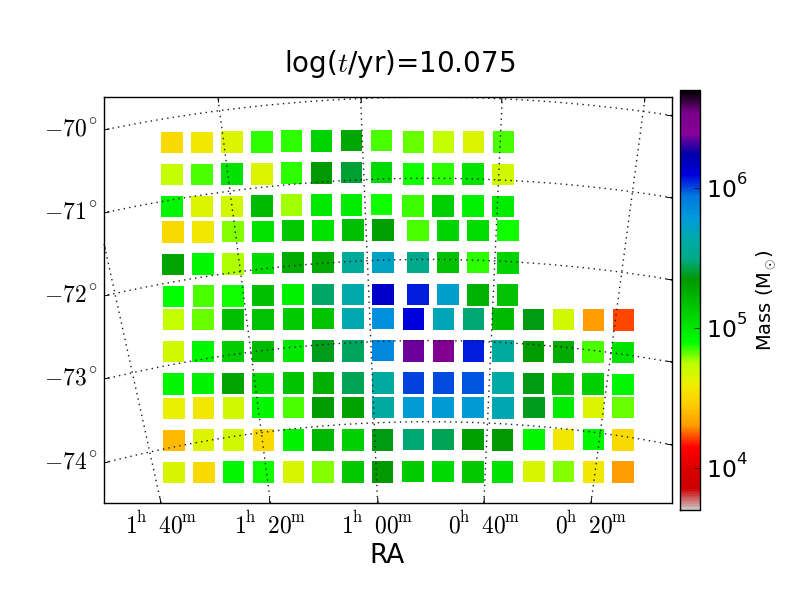}}
\resizebox{0.33\hsize}{!}{\includegraphics{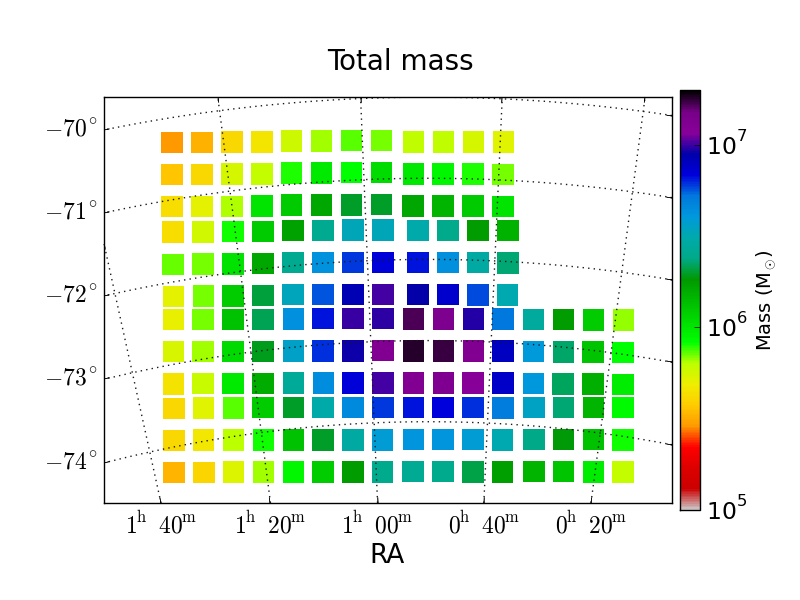}}
\hfill
\caption{The first fourteen panels show the total mass formed in each age interval considered in this work, across the SMC galaxy, expressed in $\Msun$. The bottom right-hand panel shows the inferred distribution of total stellar mass, obtained by adding the masses from all age bins.
}
\label{MASSmap}
\end{figure*}

Figure~\ref{SFRmap} displays maps of the SFR for the age bins defined in this work. They share many similarities with the SFR maps displayed by \citet{Rub15}, but cover a much wider area. Especially notable are the following points:
\begin{itemize}
\item At the youngest ages ($\log(t/\mathrm{yr})<7.8$), the SFR seems limited to areas along the SMC Bar and Wing. The Wing indeed appears like an extended blob departing from the Northern part of the Bar. It is well separated from the Southern part of the Bar by a gap in the young SFR centred at $\mathrm{RA}=1^\mathrm{h}$, which extends as far North as $\mathrm{Dec}<-73^\circ$.
\item At slightly older ages ($7.8<\log(t/\mathrm{yr})<8.5$), the separation between the Wing and the Southern part of the Bar becomes less evident. But another feature starts becoming very clear in the same age range: the Northwestern edge of the Bar becomes well delineated, as a sort of cliff line that marks a sharp reduction in the SFR towards the SMC outskirts. Stellar populations of these ages appear, globally, as a large triangle, which comprises both the Bar and the Wing, with the Northwestern edge of this triangle being delineated by this cliff line. 
\item Populations of all older ages appear with  smoother spatial distributions, becoming progressively more rounded as age increases. However, the Northwestern edge of the Bar remains still visible until ages of about $\log(t/\mathrm{yr})\lesssim9.3$, or 2~Gyr. Remarkably, the same cliff appears evident in the SFR maps from \citet{HZ04}, for ages less than 2.5~Gyr. The persistence of such a feature over such a wide age range might suggest that it is really a dynamical feature, and not just the result of recent star formation.
\item We see no evidence of the large ring-like structure found by \citet{HZ04} at an age of 2.5~Gyr. In our maps, such a structure would probably have appeared in the $\log(t/\mathrm{yr})=9.3$ age bin. What we can notice at these ages is a relatively extended plateau of uniform SFR, in the central SMC regions, without evidence of the off-centre maxima in SFR that would define a ring. For even older age bins, our SFR maxima are clearly found in the central regions, which hence more clearly excludes the presence of such a feature.
\item Populations of ages $\log(t/\mathrm{yr})\gtrsim9.7$ become quite round on the sky, finally revealing the SMC's old spheroid. Remarkably, this population is quite extended, and is not completely covered by the present data, to its Southern and Western limits.
\end{itemize}
Our results strengthen the long line of evidence for SMC populations of different ages being distributed differently \citep[e.g][]{HZ04,cignoni13}. The distinguishing feature of our analysis is the wider area covered by the VMC data and its low sensitivity to interstellar reddening. 

Integrating the SFR over each age interval, we obtain maps of the mass contribution for each age bin, which are shown in Fig.~\ref{MASSmap}. They reveal how insignificant the young SFR is compared to the mass formed in older age bins. In addition, the figure shows the distribution of total stellar mass. The latter clearly indicates that the stellar mass is concentrated around the star counts centre defined by \citet{Rub15}, rather than around the kinematic centre defined by \citet{stani04}. For comparison, both centres are indicated in the bottom right-hand panel of Fig.~\ref{SFRmap}.

\subsection{The mass assembly of the SMC}
\label{sec:mass}
 
\begin{figure}
\resizebox{0.85\hsize}{!}{\includegraphics{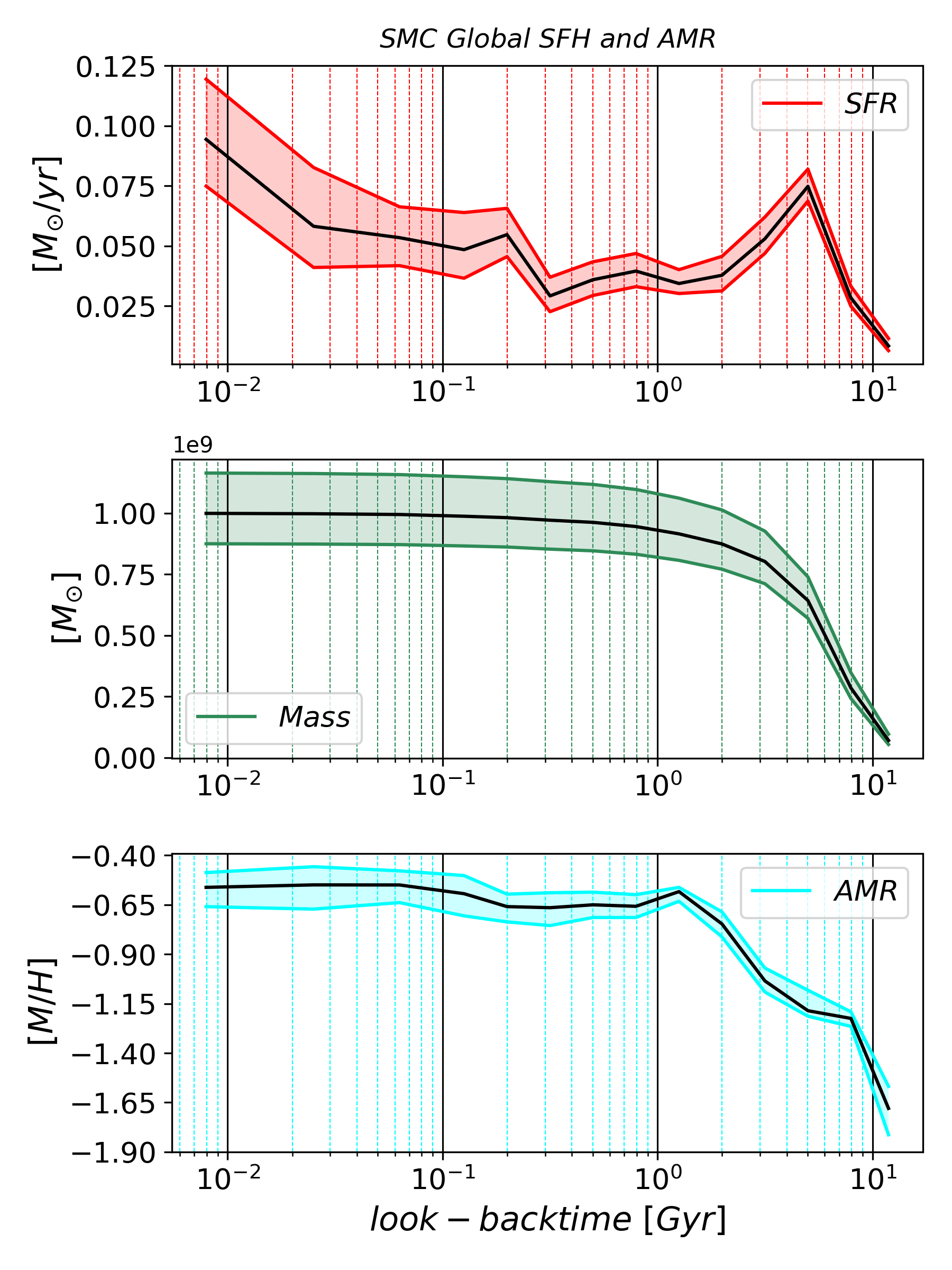}}
\caption{From top to bottom, the panels show the evolution of the global SFR, mass assembly, and AMR in the SMC, together with their confidence intervals.  }
\label{fig:gsfr}
\end{figure}

The top panel of Fig.~\ref{fig:gsfr} shows the global evolution of the SFR in the SMC, obtained from the sum of all subregions. The error bars have been simply added, hence providing an upper limit to the actual error in the sum. Integrating the SFR over the time passed since the SMC's formation $\sim\!13$~Gyr ago, we obtain the global history of stellar mass assembly in this galaxy, which is depicted in the middle panel of the figure. Of course, the interpretation of this panel as ``assembled mass'' is not strictly correct, since the stellar populations seen today are affected by a series of dynamical processes -- which moved stars far from the place of their formation, and even out of the SMC \citep[e.g.][]{olsen11} -- as well as by the reprocessing of matter inside stars -- which reassembles at younger ages part of the matter which was already assembled, and lost via stellar winds, at older ages. Therefore, the figure gives only a partial picture of how the SMC formed its present stellar mass. One can see that the SMC formed half of its stellar mass prior to an age of 6.3~Gyr. This is to be compared with the value of 8.4~Gyr found by \citet{HZ04}, and with the main SFR event at $\sim\!6$~Gyr ago seen by \citet{reza14}. Our value clearly supports the slow build-up that is typical of dwarf galaxies \citep[see][]{Weisz11,weisz14}.

The total mass of formed stars, during the entire SMC life and inside the 23.57~deg$^2$ area covered by the present work, is $5.31\pm0.05)\times10^8$~\Msun. This value depends on the assumed IMF, because a significant fraction of the inferred stellar mass is in the form of main sequence stars with masses lower than $0.8$~\Msun, which are fainter than the magnitude limit adopted in our analysis. By adopting the \texttt{PARSEC-COLIBRI} models \citep{Marigo_etal17} to describe the evolution of the stars until the end of their main nuclear burning phases -- that is, carbon burning for massive stars and the TP-AGB phase for low and intermediate-mass stars --, and the initial-to-final mass relation of white dwarfs, we can estimate that 54 per cent of this mass is still in the form of ``alive'' stars, whereas 10 per cent is in the form of stellar remnants. Their spatial distribution follows very closely the mass distribution of stars ever formed, shown in the bottom right-hand panel of Fig.~\ref{MASSmap}.

Assuming the low-mass IMF is correct, the present stellar+remnant mass, $3.4\times10^8$~\Msun, can be compared with several other SMC mass estimates in the literature, like for instance the dynamical mass of the SMC derived from its rotation curve, $2.4\times10^9$~\Msun, the total mass in cold gas, $0.7\times10^9$~\Msun\ \citep[both estimated inside a radius of 3~kpc from the kinematic centre][]{stani04}, and the total dust mass, $8.3\times10^4$~\Msun, from \citep{gordon14}. However, it is remarkable that we derive a distribution for the stellar mass which is significantly offset (by $1.3\degr$, or $\sim\!1.4$~kpc) from the kinematically derived mass, which makes any further comparison between these different masses more uncertain. Nevertheless, if we take these mass values at face value and subtract the mass presently in the form of stars, remnants, dust and cold gas from the dynamical mass, we obtain a rough estimate of $\sim1.4\times10^9$~\Msun\ for the unaccounted SMC mass. This mass could be either in the form of warm halo gas, or dark matter.

\begin{figure}
\resizebox{0.95\hsize}{!}{\includegraphics{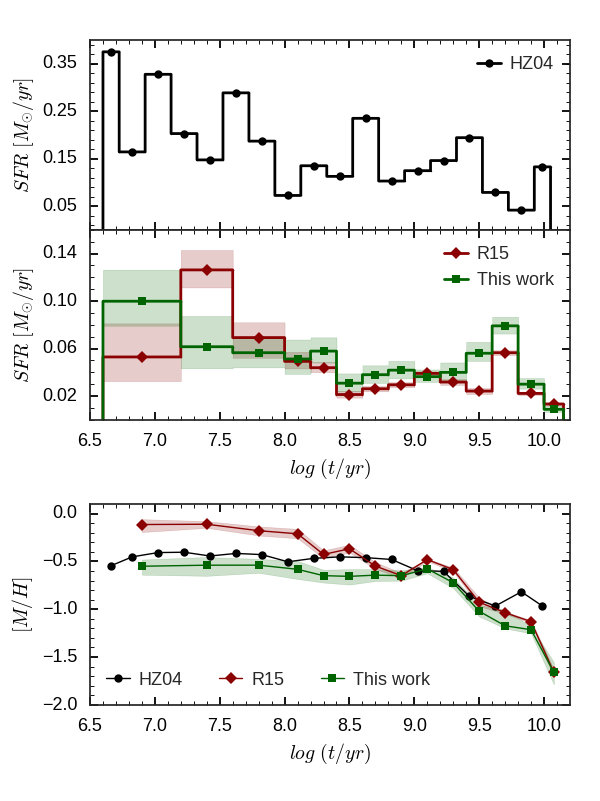}}
\caption{Comparison of the global SFR$(t)$ (top two panels) and $\mh(t)$ (bottom panel) curves obtained in this work and those from \citet{Rub15} and \citet{HZ04}. Our curves are presented as continuous lines together with shaded areas that correspond to an upper limit to the errors.}
\label{fig:hz04}
\end{figure}

At this point, it is interesting to compare our global SFR with the one derived by \citet{HZ04}, which still represents the classical work in the field, and with the analysis by \citet{Rub15}. The comparison is shown in the top two panels of Fig.~\ref{fig:hz04}, which show both the global SFR$(t)$ and the cumulative mass formed since the oldest stellar ages. The first point that can be made is that the SFR$(t)$ values by \citet{HZ04} are overall mostly in excess of ours, by factors of 2 or even more, even if their studied area is $\sim\!30$ per cent smaller. Much of this difference probably comes from the use of different IMFs,  which result into different fractions of the stellar mass being put into the form of very low-mass, low-luminosity dwarfs, which are below the detection limit of both the MCPS and the VMC survey. In this regard, \citet{HZ04} specify that they use a \citet{salpeter} IMF, without indicating the low-mass cut employed to produce a total mass of 1~\Msun. Therefore, their assumed fraction of low-mass stars is unknown. The second aspect worthy of note is that the SFR$(t)$ of \citet{HZ04} presents marked peaks which are not present in our results, or at least they are not seen at the same ages. Indeed, their SFR$(t)$ peaks at $\log(t/\mathrm{yr})$ values of $[7, 7.6, 8.6, 9.4, 10.0]$, while our peaks are less pronounced and at ages of $[6.9, 8.3, 9.7]$.  

On the other hand, comparison between the present SFR$(t)$, and the global one derived by \citet{Rub15}, in the central panel of Fig.~\ref{fig:hz04} reveals more modest differences, and mainly at the youngest ages, where the adopted stellar models have different metallicities. These differences are attributed to the joint effect of having used different areas, photometric depth, methods, IMFs, and stellar models. The most notable change is probably in the different intensity we now find for the star formation event that peaked at $\log(t/\mathrm{yr})=9.7$.

Considering the long-term evolution of the SMC, the peak at $\log(t/\mathrm{yr})=9.7$ (5~Gyr) is by far the most important feature in the SFR$(t)$. It is interesting to note that, for the ``second passage'' scenario for the interaction between the LMC and the MW \citep{patel17}, the first pericentric passage would have occurred 5~Gyr ago. If the SMC and the LMC were already a pair at the time, this passage could have triggered this major epoch of star formation.

\subsection{Metallicity evolution}
\label{sec:metal}

\begin{figure*}
\resizebox{0.33\hsize}{!}{\includegraphics{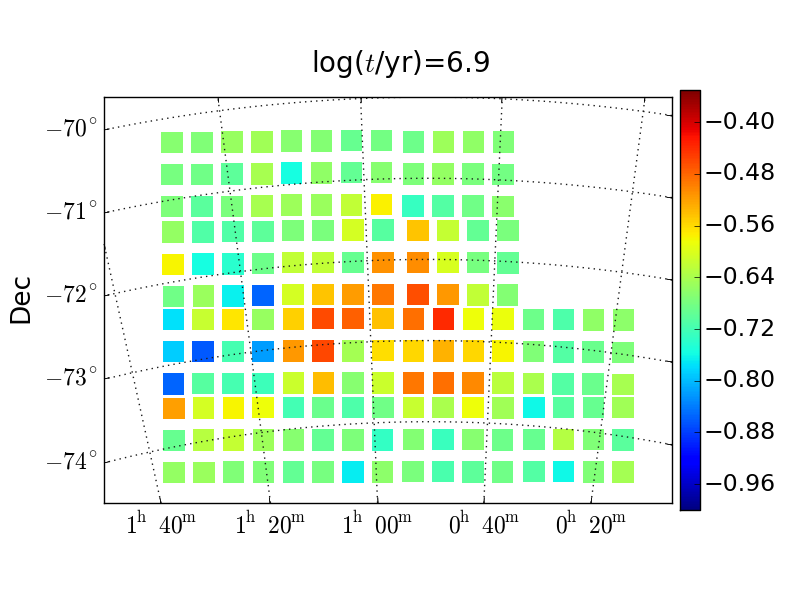}}
\resizebox{0.33\hsize}{!}{\includegraphics{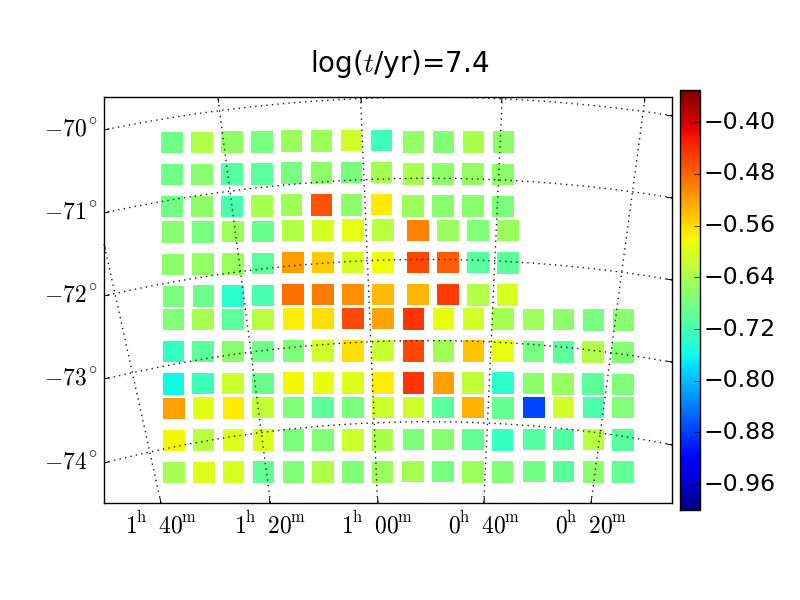}}
\resizebox{0.33\hsize}{!}{\includegraphics{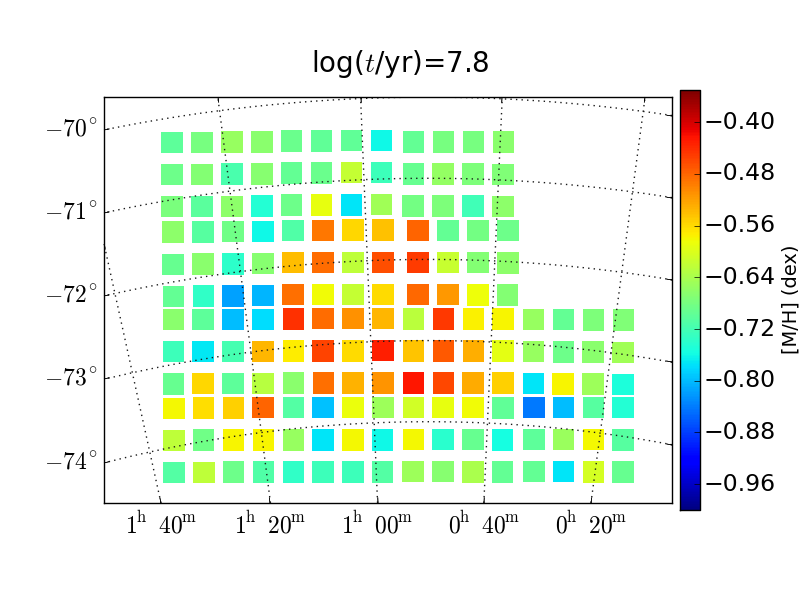}}\hfill\\
\resizebox{0.33\hsize}{!}{\includegraphics{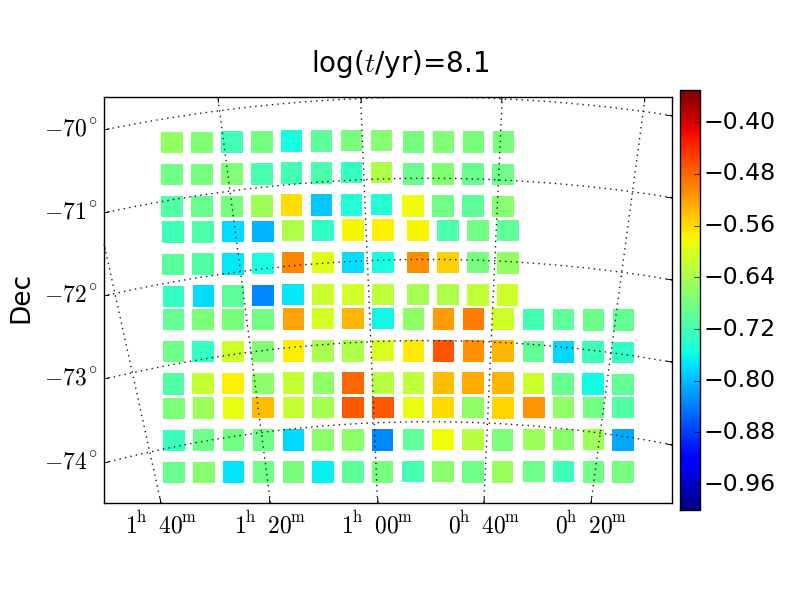}}
\resizebox{0.33\hsize}{!}{\includegraphics{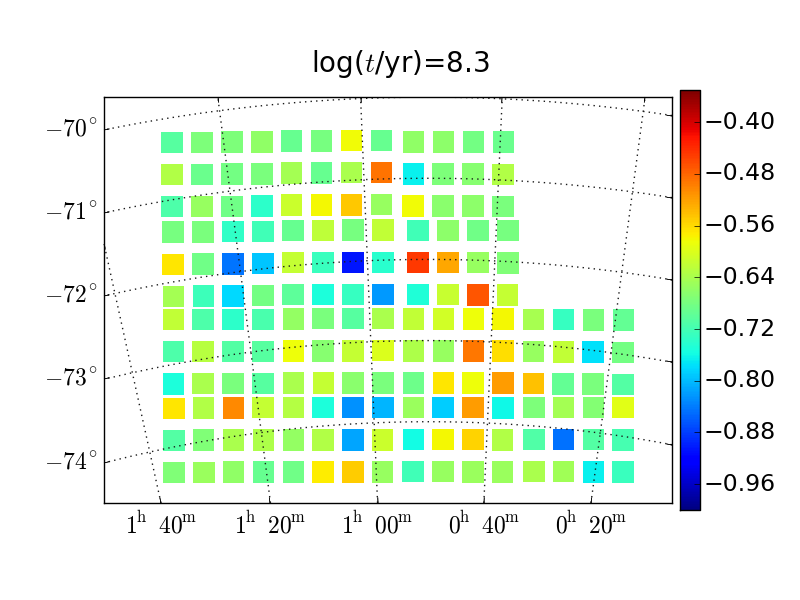}}
\resizebox{0.33\hsize}{!}{\includegraphics{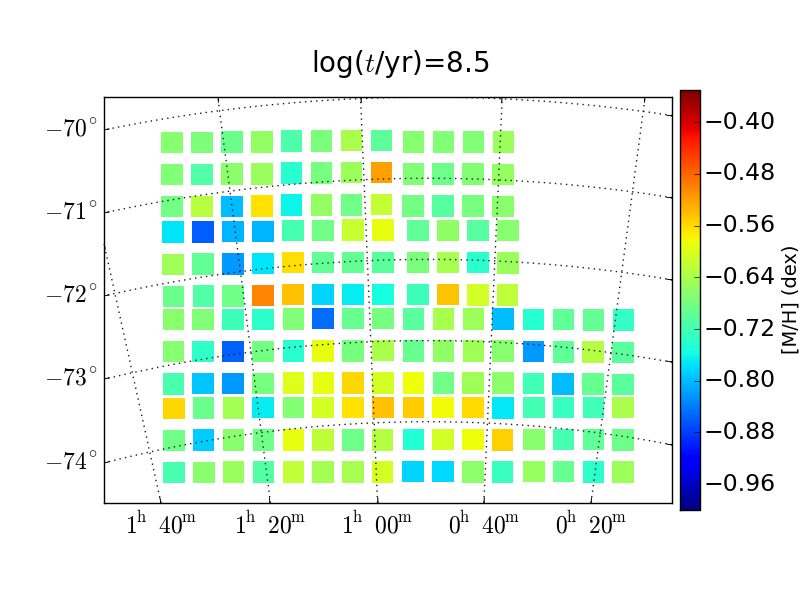}}\hfill\\
\resizebox{0.33\hsize}{!}{\includegraphics{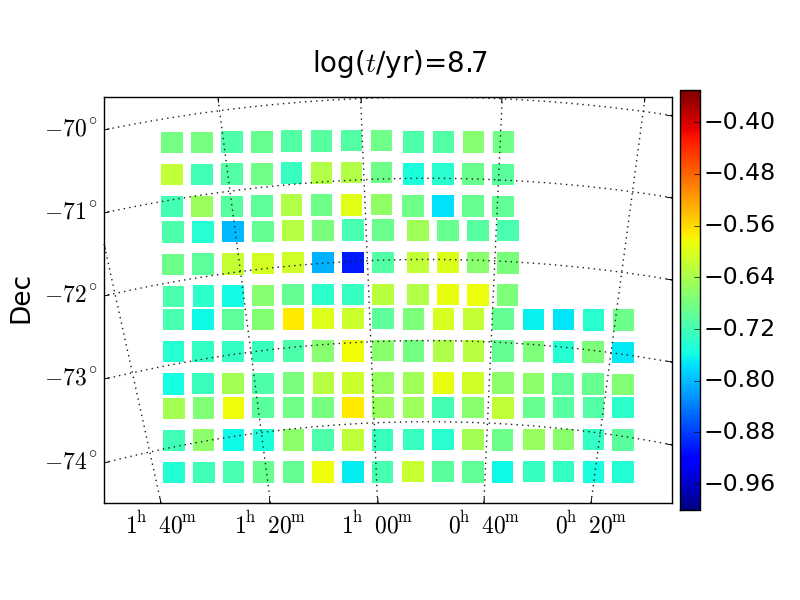}}
\resizebox{0.33\hsize}{!}{\includegraphics{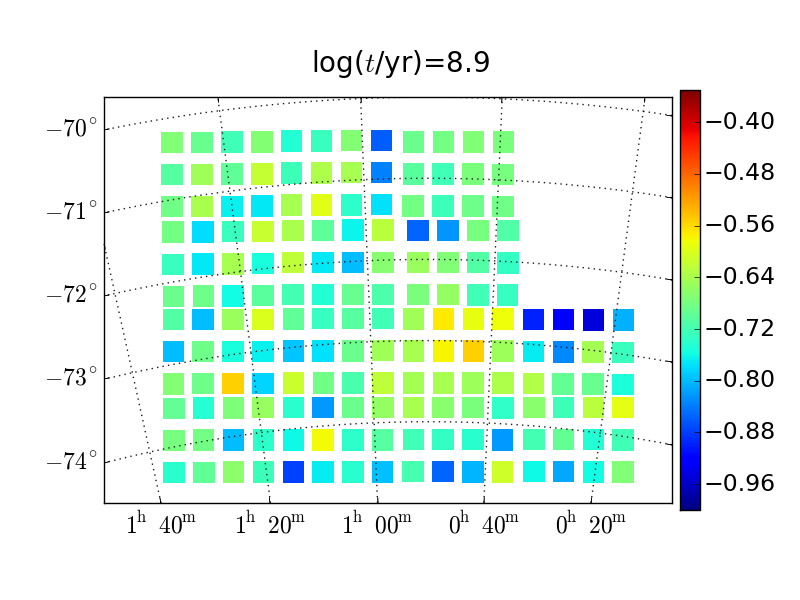}}
\resizebox{0.33\hsize}{!}{\includegraphics{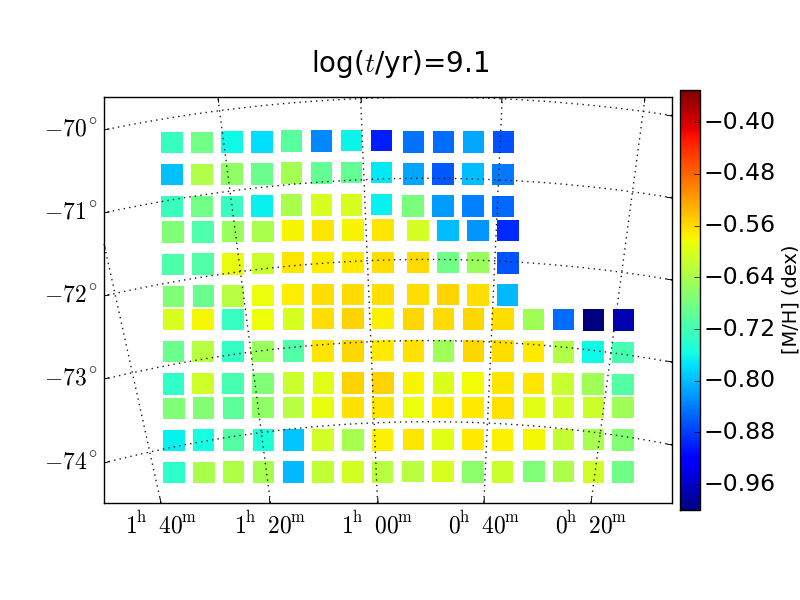}}\hfill\\
\resizebox{0.33\hsize}{!}{\includegraphics{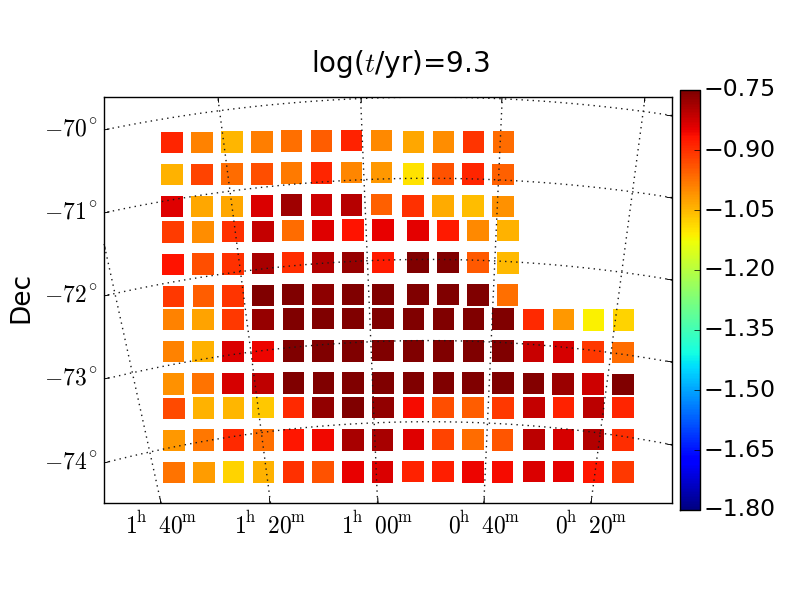}}
\resizebox{0.33\hsize}{!}{\includegraphics{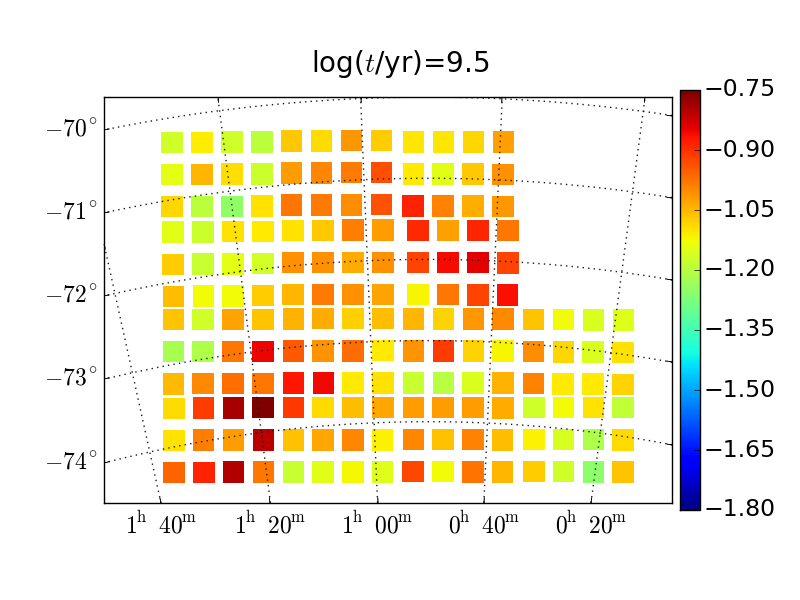}}
\resizebox{0.33\hsize}{!}{\includegraphics{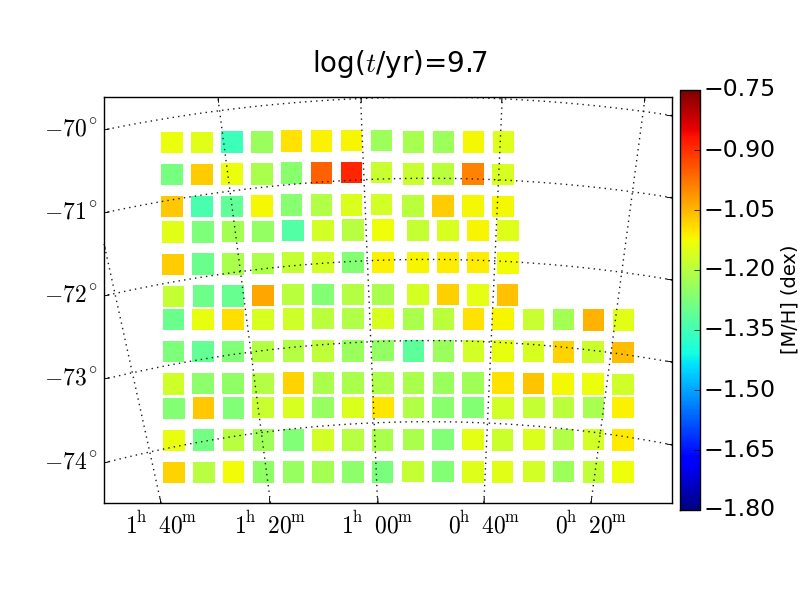}}\hfill\\
\resizebox{0.33\hsize}{!}{\includegraphics{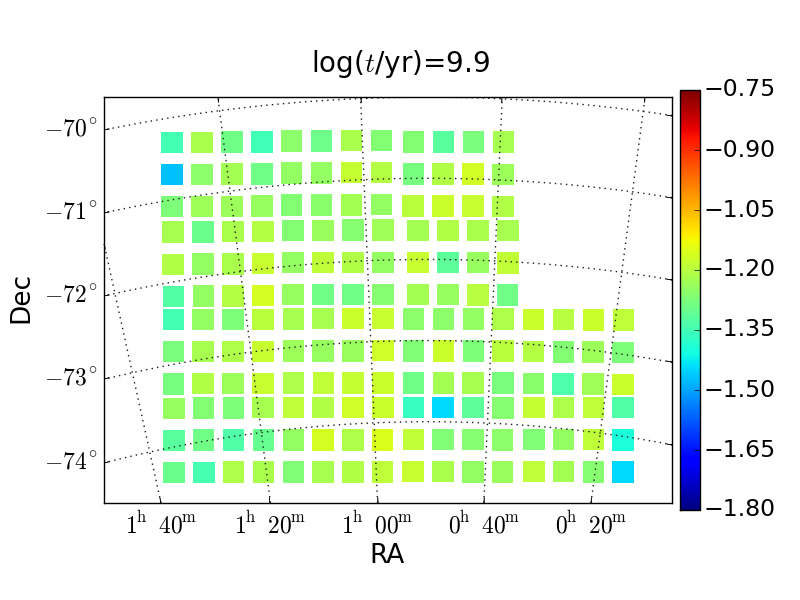}}
\resizebox{0.33\hsize}{!}{\includegraphics{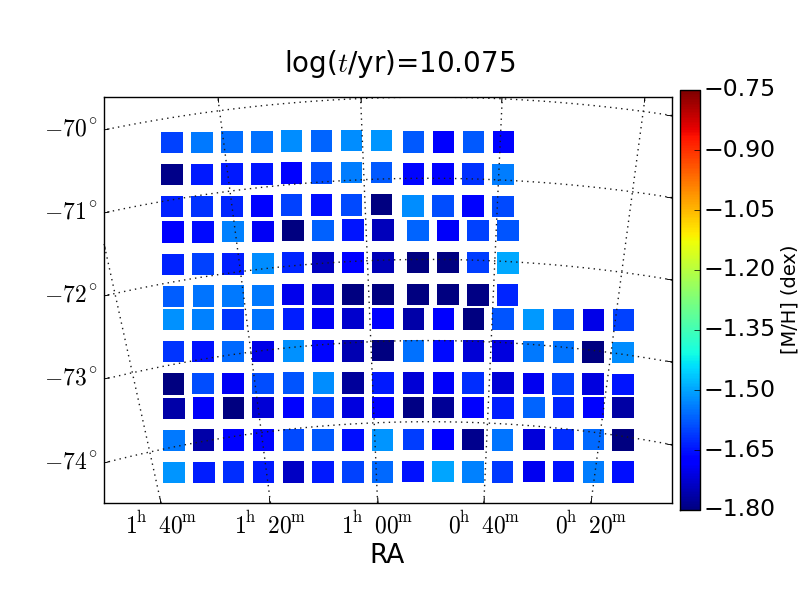}}\hfill
\caption{Maps of the mean metallicity in each age interval considered in this work. Note that the colour scale changes between the first nine panels, and the last five.}
\label{METmap}
\end{figure*}

The bottom panel of Fig.~\ref{fig:gsfr} shows the global evolution of the SMC's metallicity, averaged over all subregions (using the SFR as weight). It is evident that the youngest part of the AMR could be assumed constant from present ages up to about 130~Myr, with an average value of $\mh\sim-0.6$~dex ($Z \sim 0.0042$). Then, there is a second plateau of slightly lower metallicities ($\mh\sim-0.65$~dex or $Z \sim 0.0032$) extending up to $\sim\!1$~Gyr, which might not be significant given the uncertainties. 
Looking at the metallicity maps in Fig.~\ref{METmap}, one can see that in each of the young age intervals the spatial distribution of the average \MH\ is not homogeneous. In particular, there is evidence of a systematic increase in metallicity towards the inner Bar, where its most intense star formation is found (see Sect.~\ref{sec:sfr}). For the age interval from $\log(t/\mathrm{yr})=6.9$ to $8.1$, the typical difference in \MH\ from the outskirts to the inner Bar is about $\Delta\MH\sim0.075$~dex. Moreover, it is remarkable that the young population in the Bar seems to present a systematically higher metallicity than the young population in the Wing. It is difficult to check if these trends are real, or if they could result from the relative insensitivity of the young partial models to metallicity.
From a purely astrophysical point of view, it is possible that the young population in the Wing is more metal poor because it originates from gas that was pulled out of the SMC as a result of a dynamical interaction, and that star formation began in it at some recent epoch (at about $10^8$~yr, as indicated by the SFR maps of Fig.~\ref{SFRmap}), allowing little time for chemical self-enrichment to occur. In the centre of the SMC, instead, star formation appears to have happened more continuously over longer timescales, which might have allowed the gas to enrich in metals more than in the Wing.

Between ages from $\sim\!13$ to 1.5 Gyr ago, the AMR evolves considerably from $\MH\sim-1.6$ to $-0.65$~dex. This AMR probably reflects the main event we detect in the SFH, which is the build-up of about $80$ per cent of the entire galaxy's stellar mass between the ages of 8 and 3.5~Gyr. For the oldest ages, the spatial distributions in Fig.~\ref{METmap} do not show any significant gradient in metallicity. Such gradients appear only at $\log(t/\mathrm{yr})\leq9.3$; coincidently or not, this is also the oldest age for which the metallicity information does not originate from RGB stars, but mainly from the main sequence and core-helium burning phases.
 
Finally, the bottom panel in Fig.~\ref{fig:hz04} presents a comparison with the global AMR derived by \citet{HZ04} and by \citet{Rub15}. Most of the differences between this work and \citet{Rub15} are easily understandable, resulting mainly from the adoption, in present models, of a lower ceiling to the metallicity range at young ages (as motivated in Sect.~\ref{sec:partialmodels}). The differences with respect to \citet{HZ04} are more significant, especially at intermediate-to-old ages ($\log (t/\mathrm{yr})>9.5 $). They result, in large part, from the different choices of metallicity range: in their search of the best-fitting SFH, \citet{HZ04} use populations with just three \feh\ values, of $-1.3$, $-0.7$ and $-0.4$~dex {\em at all ages}, without allowing for very metal-poor old models as we do. These differences might be, at least in part, at the origin of the different details between our SFR$(t)$ curves -- and this is especially likely for the oldest age bin, in which \citet{HZ04} find a pronounced peak, whereas we find none (see top panels of Fig.~\ref{fig:hz04}). Of course other factors contributing to the different results include the different data, stellar models, and the assumption of a fixed distance of 60~kpc in \citet{HZ04}. Understanding the role of all these factors is beyond the scope of the current work.
 
\subsection{Typical ages of SMC populations}
\label{sec:typicalages}

\begin{figure*}
\resizebox{0.45\hsize}{!}{\includegraphics{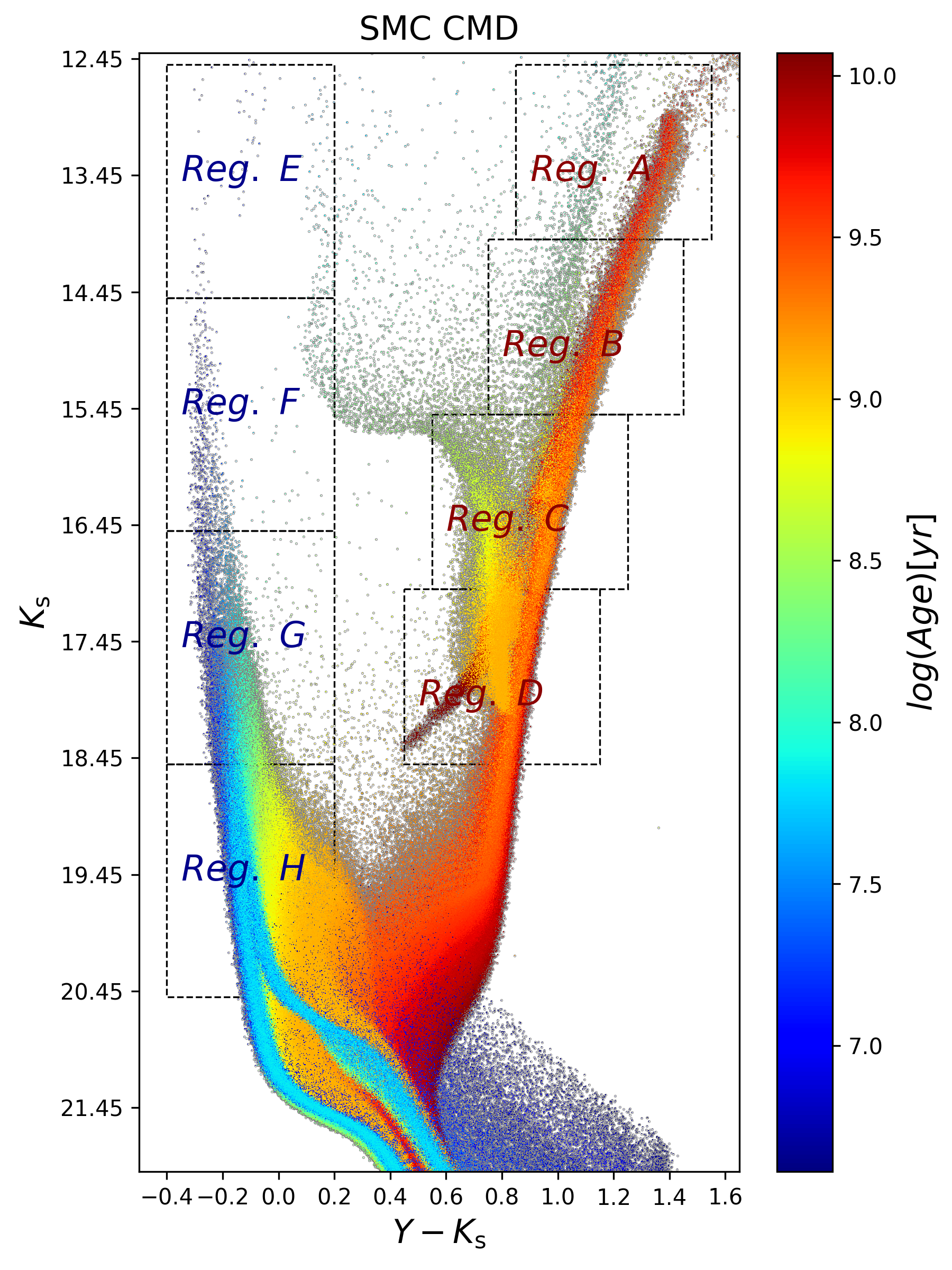}}
\resizebox{0.5\hsize}{!}{\includegraphics{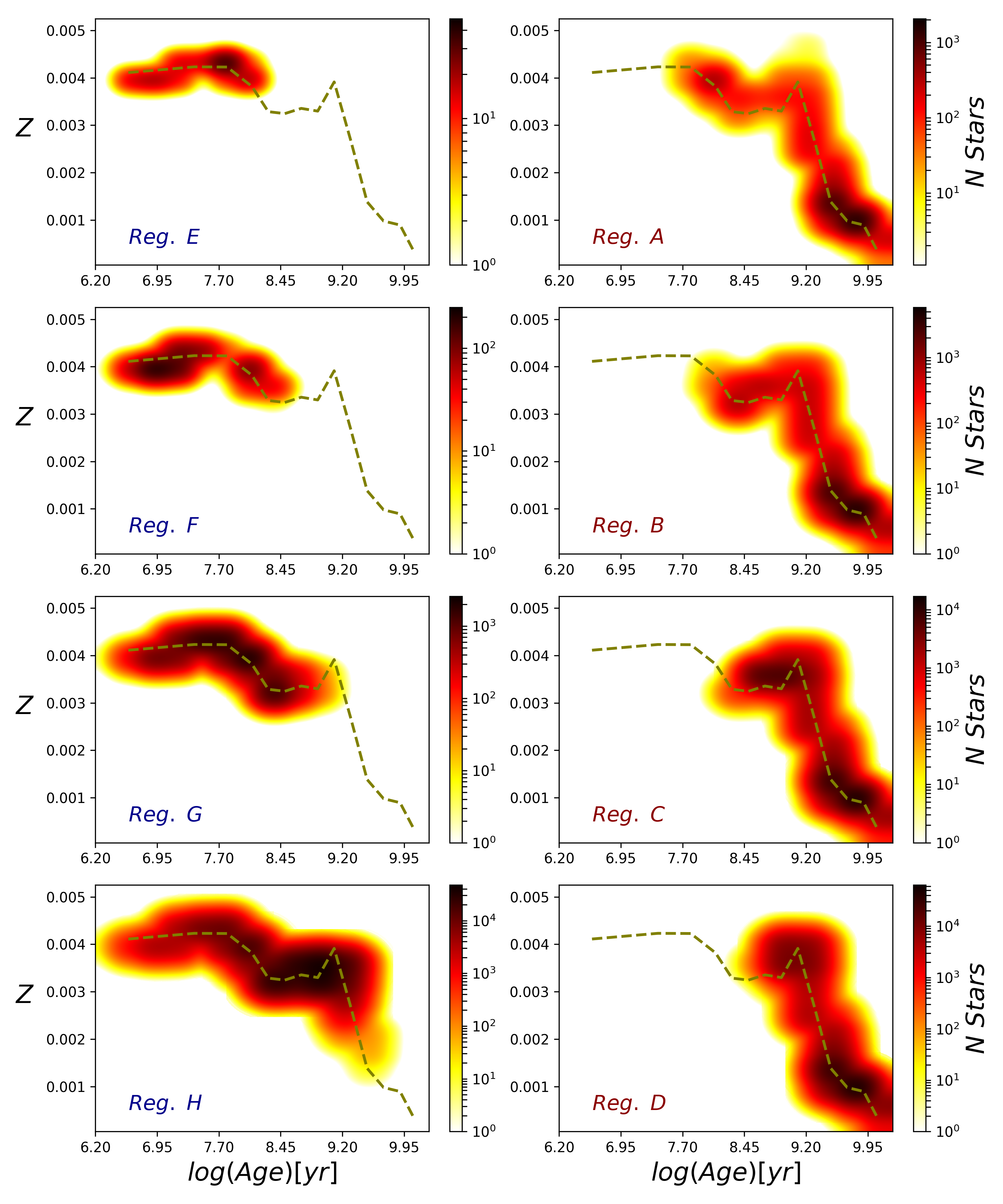}}
\caption{Overview of the stellar counts inferred for the entire SMC region investigated in this work, and their distribution in the age--metallicity plane. The {\bf left panel} shows a simulated CMD in \ks\ versus \yks, with stars colour coded by their age. The simulation includes a $30$ per cent fraction of binary stars, which are particularly evident in the bottom part of the CMD. The eight boxes are used to select different parts of the CMD. The {\bf eight right-hand panels} show the distribution of ages and metallicities for stars inside these boxes. The dashed line is the mean AMR reported in Fig.~\ref{fig:gsfr}.
}
\label{fig.smcpop}
\end{figure*}

In the coming decade, the SMC will be observed by several wide-area spectroscopic and astrometric surveys which will provide chemical abundances and kinematics for many thousands of bright stars (e.g.\ the Apache Point Observatory Galactic EVolution Experiment Southern extension, APOGEE-South, \citealt{apogeesouth}; the 4-metre Multi-Object Spectrograph Telescope, 4MOST, \citealt{4most}), as well as the bulk proper motions and parallaxes of SMC populations \citep[e.g.][]{lsst, hstpropermotions, cioni2016, gaia, vandermarel16}. These surveys promise to provide extremely tight constraints on the SMC's evolutionary history. The interpretation of such data, however, might not be straightforward since different stellar samples might represent very different age (and hence metallicity) distributions. To aid the interpretation of future spectroscopic and astrometric data, Fig.~\ref{fig.smcpop} presents an overview of the age distributions implied by our SFH analysis, for the entire SMC galaxy, and for a few selected regions of the near-infrared CMD. The left panel in this figure simply shows the CMD for the entire SMC, simulated with the TRILEGAL code using the best-fitting SFH for each subregion, but placing each of these subregions at the same reference value of true distance modulus and extinction, namely at $\dmo=18.90$~mag and $\av=0.35$~mag. In other words, it represents the reconstructed SMC populations that would have been observed under ideal conditions (i.e.\ no photometric errors, no zero-point uncertainties, no spread in distance and extinction, and no foreground galaxy).

The right-hand panels in Fig.~\ref{fig.smcpop} illustrate the distribution of ages and metallicities for the regions of the CMD outlined in the left panel. For the boxes placed along the main sequence (from E to H), it is no surprise that the fainter boxes sample progressively older (and more metal-poor) populations. Even the faintest bins along the main sequence do always contain a sizeable fraction of young stars.
Indeed, a good separation of the population ages is present only in the subgiant region of the CMD. Along the red giant branch, the boxes show that the entire sequence contains a similar distribution of ages and metallicities as regards the intermediate-to-old populations (that is, for all ages older than $\sim2\times10^9$~yr).
However, younger populations appear in very different proportions: for instance, the brightest box A contains a significant blob at $10^8$~yr, which corresponds to the RSG sequence already illustrated in Fig.~\ref{fig:metlim}; C and D instead contain a prominent population of ages $\sim2\times10^9$~yr, caused by red clump stars \citep[see][]{girardi16}, which in the case of box C extend to slightly younger ages.   

Since Fig.~\ref{fig.smcpop} represents the final mean model of the SMC population, it can be redone for any sub-area, photometric system and depth employed by present-day surveys, starting from the SFH tables provided in this paper.

\section{Concluding remarks and future work}
\label{sec:conclu}

In this work, we present an analysis of VMC data comprising the main body of the SMC galaxy. Re-construction of the observed CMDs allows us to derive the best-fitting SFH, distance and extinction across the Bar and Wing regions. Our analysis gives a broad picture of how the stellar ages, assembled mass, and mean metallicities are distributed spatially, both across the sky and in terms of their mean distance. The use of near-infrared filters allows us to reduce the impact of the differential and internal extinction, which is the main complicating factor for this kind of analysis. This picture updates the classic one provided by \citet{HZ04} for the inner 18~deg$^2$ of the SMC, and supersedes the one provided by \citet{Rub15} for a smaller and non-contiguous area. Our results are significantly different from those of \citet{HZ04} in many respects. Since both works include the bulk of the SMC stellar populations, the differences have to be ascribed to the different data, methods, and stellar models used in both cases, rather than to the present $\sim\!30$ per cent increase in the area studied.

The a posteriori analysis of the results reveals that at the maximum distance from the SMC centre included in the present data (i.e.\ $\sim\!4^\circ$ in the northeastern direction), the density of the old SMC population still decreases outwards. This implies that our present VMC data do not comprise the entire old spheroid of the galaxy -- which is not surprising given the power-law decrease in density, up to large distances from the SMC centre, inferred by \citet{noel09}, and which seems supported by our own stellar density maps (see bottom right-hand panel in Fig.~\ref{MASSmap}). The coverage of the SMC area will be improved in the future, since the VMC footprint includes additional sets of tiles south and east of the presently studied area, as well as a sequence of tiles across the Magellanic Bridge \citep[see][]{cioni11}. However, the outermost SMC areas are somewhat problematic for our SFH analysis, owing to their small SMC stellar densities and a large fraction of the observed stars belonging to the Milky Way foreground -- although the same data is probably very useful for simpler analyses based on the spatial variations of the star counts within larger CMD boxes, like for instance those presented by \citet{noel09}, \citet{bagheri13} and \citet{skowron14}. We recall that the outer parts of the SMC spheroid have been more completely mapped using RR~Lyrae \citep{jacy17, Muraveva18}, and are targeted in a more systematic way in optical passbands, by the Optical Gravitational Lensing Experiment IV \citep[OGLE-IV;][]{ogleiv}, Survey of the MAgellanic Stellar History \citep[SMASH;][]{smash}, the SMC in Time: Evolution of a Prototype interacting dwarf galaxy \citep[STEP;][]{step}, Magellanic Satellites Survey \citep[MagLites;][]{torrealba18}, and ``Yes, Magellanic Clouds Again'' (YMCA; Ripepi et al, in prep.). They will also be further and more extensively spatially mapped by Gaia \citep[see][]{gaia, belokurov17}. 

Our results will provide key constraints on the theoretical models of the SMC evolution and of its interaction with the LMC and the Milky Way. For instance, we have revised epochs for the main periods of star formation in the SMC, new results indicating how the SMC metallicity evolved during the past two LMC/SMC interactions and how metals were radially mixed over the last 0.2 Gyr. We provide maps for the mass distribution of stars of different ages, and a total mass of stars ever formed, which can be directly compared to the results of N-body and hydrodynamical models for the SMC evolution. Our results for the variation of the mean distance across the SMC, being more sensitive to the intermediate-to-old populations that dominate the near-infrared CMDs, can be used either in addition or as alternative to results based on other distance indicators. Together with reasonable assumptions about the mass fraction locked in stars and stellar remnants, and the IMF, we estimate the present mass in stellar objects, that ultimately may help to constrain dynamical models and the dark matter halo mass of the SMC. We also have new clues of points where stellar models at sub-solar metallicities can be improved, in order to better reproduce the observed star counts and Hess diagrams. Such information is hard to obtain from smaller-area surveys, or from optical surveys where the interstellar extinction represents a major source of scatter in the photometry.

\subsection*{Acknowledgements}
This research is supported by the European Research Council (ERC) consolidator grant funding scheme for SR, GP, PM and AN (project STARKEY, G.A. n.~615604), as well as for MRLC (project INTERCLOUDS, G.A. n.~682115). RdG acknowledges funding from the National Natural Science Foundation of China (grants U1631102 and 11633005). He also received support from the National Key Research and Development Program of China through grant 2017YFA0402702. 

We thank CASU and WFAU for providing calibrated data products under the support of the Science and Technology Facility Council (STFC) in the UK. We thank Karl Gordon for kindly sharing his extinction curve codes. This research is based on observations made with VISTA at ESO under programme ID 179.B-2003.


\appendix
\section{Results for all subregions}
\label{sec:catalog}
 
Table~\ref{tabellone} presents a summary of the quantities relevant for the analysis presented in this paper, and potentially useful for independent analyses of SMC data. They include the true distance moduli, extinction values, total SFR and mean \mh\ in each age bin, for all tiles and subregions, together with their $1\sigma$ (68 per cent) confidence intervals.

\setlength{\tabcolsep}{2pt}

\begin{landscape}
\begin{table}
\caption{Sample table summarising the results of the SFH analysis. A computer-readable version, complete with all age bins and subregions, can be downloaded from \url{http://stev.oapd.inaf.it/VMC_SFHdata}. 
}
\label{tabellone}
\begin{scriptsize}
\begin{tabular}{ccc|ccccccc|ccccccc|ccccccc|ccccccccc}
\hline\hline
tile & subr. & RA & DEC & 
$(m\mathrm{-}M)_0^{JK\mathrm{s}}$ & $\sigma$ & 
$A_V^{JK\mathrm{s}}$ & $\sigma$ & 
$(m\mathrm{-}M)_{0,\mathrm{min}}^{JK\mathrm{s}}$ & $A_{V,\mathrm{min}}^{JK\mathrm{s}}$ & 
$\chi_{\mathrm{min}}^{2,JK\mathrm{s}}$ & 
$(m\mathrm{-}M)_0^{YK\mathrm{s}}$ & $\sigma$ & 
$A_V^{YK\mathrm{s}}$ & $\sigma$ & 
$(m\mathrm{-}M)_{0,\mathrm{min}}^{YK\mathrm{s}}$ & $A_{V,\mathrm{min}}^{YK\mathrm{s}}$ & 
$\chi_{\mathrm{min}}^{2,YK\mathrm{s}}$ & 
$\log t_1$ & SFR$_1$ & 
SFR$_1^\mathrm{l}$ & SFR$_1^\mathrm{u}$ & 
$\mh_1$ & $\mh_1^\mathrm{l}$ & $\mh_1^\mathrm{u}$ & 
$\log t_2$ & SFR$_2$ & 
SFR$_2^\mathrm{l}$ & SFR$_2^\mathrm{u}$ & 
$\mh_2$ & $\mh_2^\mathrm{l}$ & $\mh_2^\mathrm{u}$ & 
$\cdots$ \\
SMC   &       & \degr & \degr & mag & mag & mag & mag & mag & mag &  & mag & mag & mag & mag & mag & mag & &
     [yr] & $10^{-3}$ & $10^{-3}$ & $10^{-3}$ & dex & dex & dex & 
     [yr] & $10^{-3}$ & $10^{-3}$ & $10^{-3}$ & dex & dex & dex & 
     $\cdots$ \\
   &       &  &  &  &  &  &  &  &  &  &  &  &  &  &  &  & &
      & $\Msun/\mathrm{yr}$ & $\Msun/\mathrm{yr}$ & $\Msun/\mathrm{yr}$ &  &  &  & 
      & $\Msun/\mathrm{yr}$ & $\Msun/\mathrm{yr}$ & $\Msun/\mathrm{yr}$ &  &  &  & 
     $\cdots$ \\
     &       & (1) & (1) & (2) & (2) & (2) & (2) & (3) & (3) & (4) & (2) & (2) & (2) & (2) & (3) & (3) & (4) &
     (5) & (6) & (6) & (6) & (7) & (7) & (7) & 
     (5) & (6) & (6) & (6) & (7) & (7) & (7) & 
     $\cdots$ \\
\hline
3\_2 & 1 & 7.78422 & -74.55999 & 19.022 & 0.067 & 0.432 & 0.158 & 19.025 & 0.425 & 1.330 & 19.000 & 0.115 & 0.321 & 0.094 & 18.975 & 0.325 & 1.108 & 6.9 & 0.045 & 0.011 & 0.105 & -0.706 & -0.966 & -0.471 & 7.4 & 0.018 & 0.001 & 0.060 & -0.684 & -0.936 & -0.473 & $\cdots$ \\
3\_2 & 2 & 6.42938 & -74.52585 & 18.970 & 0.093 & 0.433 & 0.097 & 18.975 & 0.425 & 1.580 & 19.050 & 0.069 & 0.266 & 0.138 & 19.050 & 0.250 & 1.581 & 6.9 & 0.211 & 0.173 & 0.388 & -0.761 & -0.959 & -0.540 & 7.4 & 0.128 & 0.059 & 0.223 & -0.700 & -0.897 & -0.486 & $\cdots$ \\
3\_2 & 3 & 5.08099 & -74.48350 & 18.989 & 0.124 & 0.447 & 0.117 & 18.975 & 0.475 & 1.493 & 18.984 & 0.139 & 0.331 & 0.208 & 19.025 & 0.350 & 1.409 & 6.9 & 0.010 & 0.001 & 0.059 & -0.671 & -0.880 & -0.433 & 7.4 & 0.049 & 0.018 & 0.094 & -0.664 & -0.968 & -0.442 & $\cdots$ \\
3\_2 & 4 & 3.74048 & -74.43303 & 18.972 & 0.070 & 0.375 & 0.092 & 18.975 & 0.375 & 1.325 & 18.993 & 0.138 & 0.252 & 0.181 & 19.050 & 0.200 & 1.305 & 6.9 & 0.002 & 0.000 & 0.022 & -0.643 & -0.854 & -0.494 & 7.4 & 0.006 & 0.001 & 0.019 & -0.695 & -0.875 & -0.491 & $\cdots$ \\
3\_2 & 5 & 7.90438 & -74.16571 & 18.990 & 0.081 & 0.452 & 0.093 & 19.000 & 0.450 & 2.064 & 19.018 & 0.077 & 0.300 & 0.057 & 19.050 & 0.275 & 1.796 & 6.9 & 0.018 & 0.001 & 0.111 & -0.692 & -0.920 & -0.458 & 7.4 & 0.080 & 0.003 & 0.178 & -0.703 & -0.934 & -0.458 & $\cdots$ \\
3\_2 & 6 & 6.58207 & -74.13238 & 19.000 & 0.072 & 0.462 & 0.091 & 19.025 & 0.450 & 1.679 & 18.980 & 0.130 & 0.302 & 0.185 & 18.975 & 0.325 & 1.938 & 6.9 & 0.048 & 0.001 & 0.162 & -0.630 & -0.886 & -0.440 & 7.4 & 0.180 & 0.075 & 0.284 & -0.706 & -0.899 & -0.490 & $\cdots$ \\
3\_2 & 7 & 5.26574 & -74.09106 & 19.003 & 0.060 & 0.448 & 0.114 & 19.000 & 0.450 & 1.330 & 19.014 & 0.087 & 0.342 & 0.071 & 19.025 & 0.325 & 1.464 & 6.9 & 0.015 & 0.001 & 0.071 & -0.672 & -0.892 & -0.444 & 7.4 & 0.035 & 0.001 & 0.081 & -0.625 & -0.950 & -0.422 & $\cdots$ \\
3\_2 & 8 & 3.95674 & -74.04181 & 18.999 & 0.060 & 0.430 & 0.151 & 18.975 & 0.450 & 1.439 & 18.981 & 0.106 & 0.289 & 0.117 & 18.950 & 0.300 & 1.507 & 6.9 & 0.010 & 0.000 & 0.032 & -0.702 & -0.890 & -0.495 & 7.4 & 0.004 & 0.001 & 0.029 & -0.705 & -0.939 & -0.474 & $\cdots$ \\
3\_2 & 9 & 8.01885 & -73.77137 & 18.982 & 0.052 & 0.429 & 0.062 & 18.975 & 0.425 & 2.687 & 18.967 & 0.071 & 0.333 & 0.035 & 18.950 & 0.350 & 2.646 & 6.9 & 0.155 & 0.025 & 0.475 & -0.761 & -0.979 & -0.524 & 7.4 & 0.549 & 0.304 & 0.741 & -0.875 & -0.982 & -0.662 & $\cdots$ \\
3\_2 & 10 & 6.72755 & -73.73881 & 19.002 & 0.072 & 0.414 & 0.097 & 19.000 & 0.425 & 2.023 & 19.021 & 0.093 & 0.296 & 0.074 & 19.025 & 0.325 & 2.084 & 6.9 & 0.063 & 0.003 & 0.156 & -0.703 & -0.977 & -0.457 & 7.4 & 0.073 & 0.008 & 0.163 & -0.608 & -0.888 & -0.416 & $\cdots$ \\
3\_2 & 11 & 5.44182 & -73.69846 & 19.015 & 0.080 & 0.417 & 0.116 & 19.000 & 0.425 & 1.601 & 19.008 & 0.056 & 0.308 & 0.056 & 19.000 & 0.325 & 1.489 & 6.9 & 0.015 & 0.001 & 0.076 & -0.691 & -0.919 & -0.446 & 7.4 & 0.027 & 0.004 & 0.067 & -0.713 & -0.979 & -0.453 & $\cdots$ \\
3\_2 & 12 & 4.16290 & -73.65037 & 19.013 & 0.078 & 0.346 & 0.110 & 19.025 & 0.350 & 1.392 & 19.018 & 0.034 & 0.254 & 0.093 & 19.025 & 0.250 & 1.334 & 6.9 & 0.033 & 0.003 & 0.073 & -0.647 & -0.959 & -0.447 & 7.4 & 0.008 & 0.001 & 0.038 & -0.672 & -0.881 & -0.479 & $\cdots$ \\
3\_3 & 1 & 13.30525 & -74.61343 & 18.970 & 0.113 & 0.400 & 0.097 & 18.975 & 0.400 & 1.276 & 18.997 & 0.088 & 0.556 & 0.200 & 19.000 & 0.550 & 1.219 & 6.9 & 0.003 & 0.000 & 0.027 & -0.676 & -0.944 & -0.453 & 7.4 & 0.003 & 0.000 & 0.024 & -0.644 & -0.869 & -0.444 & $\cdots$ \\
3\_3 & 2 & 11.93819 & -74.61484 & 18.915 & 0.075 & 0.463 & 0.081 & 18.900 & 0.475 & 1.732 & 18.917 & 0.071 & 0.600 & 0.053 & 18.925 & 0.600 & 1.399 & 6.9 & 0.021 & 0.001 & 0.052 & -0.714 & -0.922 & -0.438 & 7.4 & 0.007 & 0.001 & 0.031 & -0.678 & -0.926 & -0.446 & $\cdots$ \\
3\_3 & 3 & 10.57154 & -74.6079 & 19.069 & 0.059 & 0.228 & 0.075 & 19.075 & 0.200 & 1.523 & 19.014 & 0.072 & 0.573 & 0.080 & 19.000 & 0.575 & 1.344 & 6.9 & 0.038 & 0.008 & 0.082 & -0.699 & -0.951 & -0.475 & 7.4 & 0.009 & 0.001 & 0.039 & -0.648 & -0.909 & -0.437 & $\cdots$ \\
3\_3 & 4 & 9.20687 & -74.59263 & 19.015 & 0.037 & 0.350 & 0.067 & 19.025 & 0.325 & 1.442 & 19.018 & 0.070 & 0.515 & 0.076 & 19.025 & 0.500 & 1.245 & 6.9 & 0.024 & 0.001 & 0.060 & -0.683 & -0.915 & -0.446 & 7.4 & 0.009 & 0.001 & 0.044 & -0.670 & -0.916 & -0.446 & $\cdots$ \\
3\_3 & 5 & 13.28299 & -74.21787 & 18.929 & 0.062 & 0.407 & 0.052 & 18.925 & 0.400 & 1.893 & 18.995 & 0.056 & 0.565 & 0.037 & 18.975 & 0.550 & 1.880 & 6.9 & 0.022 & 0.001 & 0.066 & -0.669 & -0.945 & -0.434 & 7.4 & 0.026 & 0.001 & 0.091 & -0.673 & -0.921 & -0.442 & $\cdots$ \\
3\_3 & 6 & 11.94933 & -74.21922 & 19.020 & 0.061 & 0.367 & 0.083 & 19.025 & 0.350 & 2.185 & 19.000 & 0.047 & 0.600 & 0.067 & 19.025 & 0.575 & 1.629 & 6.9 & 0.042 & 0.005 & 0.100 & -0.727 & -0.974 & -0.473 & 7.4 & 0.034 & 0.002 & 0.121 & -0.666 & -0.930 & -0.435 & $\cdots$ \\
3\_3 & 7 & 10.61606 & -74.21245 & 19.040 & 0.037 & 0.385 & 0.076 & 19.050 & 0.375 & 2.276 & 19.021 & 0.052 & 0.625 & 0.061 & 19.025 & 0.625 & 1.640 & 6.9 & 0.090 & 0.007 & 0.187 & -0.667 & -0.946 & -0.429 & 7.4 & 0.104 & 0.019 & 0.207 & -0.693 & -0.927 & -0.462 & $\cdots$ \\
3\_3 & 8 & 9.28461 & -74.19757 & 19.025 & 0.045 & 0.325 & 0.061 & 19.025 & 0.325 & 1.919 & 19.006 & 0.032 & 0.537 & 0.084 & 19.000 & 0.550 & 1.587 & 6.9 & 0.037 & 0.004 & 0.092 & -0.687 & -0.953 & -0.440 & 7.4 & 0.097 & 0.010 & 0.206 & -0.729 & -0.937 & -0.470 & $\cdots$ \\
3\_3 & 9 & 13.2618 & -73.82229 & 19.000 & 0.061 & 0.329 & 0.052 & 19.000 & 0.325 & 2.570 & 19.015 & 0.037 & 0.640 & 0.037 & 19.000 & 0.625 & 2.907 & 6.9 & 0.107 & 0.006 & 0.300 & -0.615 & -0.883 & -0.408 & 7.4 & 0.237 & 0.077 & 0.405 & -0.610 & -0.800 & -0.455 & $\cdots$ \\
3\_3 & 10 & 11.95994 & -73.82360 & 19.017 & 0.035 & 0.367 & 0.035 & 19.025 & 0.350 & 2.826 & 19.025 & 0.035 & 0.662 & 0.037 & 19.025 & 0.675 & 3.354 & 6.9 & 0.493 & 0.252 & 0.716 & -0.637 & -0.867 & -0.495 & 7.4 & 0.450 & 0.220 & 0.664 & -0.700 & -0.915 & -0.474 & $\cdots$ \\
3\_3 & 11 & 10.65845 & -73.81699 & 19.015 & 0.037 & 0.405 & 0.056 & 19.025 & 0.400 & 2.732 & 19.017 & 0.035 & 0.692 & 0.035 & 19.025 & 0.700 & 3.037 & 6.9 & 0.284 & 0.140 & 0.499 & -0.584 & -0.852 & -0.450 & 7.4 & 0.627 & 0.368 & 0.884 & -0.531 & -0.736 & -0.423 & $\cdots$ \\
3\_3 & 12 & 9.35865 & -73.80247 & 19.033 & 0.035 & 0.283 & 0.071 & 19.025 & 0.300 & 2.321 & 19.030 & 0.056 & 0.585 & 0.037 & 19.025 & 0.575 & 2.481 & 6.9 & 0.350 & 0.169 & 0.540 & -0.647 & -0.870 & -0.480 & 7.4 & 0.409 & 0.179 & 0.739 & -0.692 & -0.881 & -0.489 & $\cdots$ \\
$\vdots$ &$\vdots$ &$\vdots$ &$\vdots$ &$\vdots$ &$\vdots$ &$\vdots$ &$\vdots$ &$\vdots$ &$\vdots$ &$\vdots$ &$\vdots$ &$\vdots$ &$\vdots$ &$\vdots$ &$\vdots$ &$\vdots$ &$\vdots$ &$\vdots$ &$\vdots$ &$\vdots$ &$\vdots$ &$\vdots$ &$\vdots$ &$\vdots$ &$\vdots$ &$\vdots$ &$\vdots$ &$\vdots$ &$\vdots$ &$\vdots$ &$\vdots$ &$\ddots$ \\
\hline
\end{tabular}
\\
Table notes:\\
(1) J2000 coordinates for the centre of subregions. Each subregion covers an area of about 0.143~deg$^2$, but for subregions 1 where the area is $\sim\!37$\% smaller. \\
(2) True distance modulus and extinction inferred from the polynomial fit (Sect.~\ref{sec:decoupling}), with $1\sigma$ error, either in the $J\ks$ or $Y\ks$ CMD. \\
(3) True distance modulus and $\av$ from the minimum value of $\chi^2$, and (4) its corresponding value, \chisqmin, either in the $J\ks$ or $Y\ks$ CMD. \\
(5) Each one of the 14 age bins defined in Sect.~\ref{sec:partialmodels}. Each age bin is followed by (6) the total SFR for that bin, together with lower and upper limits (68\% confidence level), and (7) the mean \mh\ for that bin, again with lower and upper limits. These latter quantities are averaged from the $Y\ks$ and $J\ks$ CMD solutions.
\end{scriptsize}
\end{table}
\end{landscape}

\section{The v1.5 calibration of VISTA data}
\label{app-calib}

As mentioned in Sect.~\ref{sec:decoupling}, while our analysis of VISTA v1.3 data was ongoing, a recalibration of VISTA photometry was made available by \citet{gonzalez17}. It is incorporated into the newest version (v1.5) of VDFS which is currently being applied to all observations with VISTA prior to 2017 (including those used in this paper). 
Here, we simply check how the new calibration would have affected the colour offsets found between best-fitting models and data in the $Y\ks$ and $J\ks$ CMDs, that are the major motivation for decoupling the two CMDs in the present analysis (Sect.~\ref{sec:decoupling}).

According to Appendix C2 in \citet{gonzalez17}, the mean magnitude differences between the VISTA v1.3 and v1.5 data are 
\begin{eqnarray}
Y_{1.3}-Y_{1.5} &=& 0.018\pm0.004 \mathrm{~mag}\\
J_{1.3}-J_{1.5} &=& -0.0200\pm0.0008 \mathrm{~mag} \\
\ks_{1.3}-\ks_{1.5} &=& 0.0106\pm0.0007 \mathrm{~mag}
\end{eqnarray}
In addition, we have to consider that \citep{gonzalez17} derive the value to use to convert VISTA $Y$, by imposing that the mean colours of observed A0V stars are zero, on average.
This latter aspect is already included in our model realisation of Vega magnitudes, which strictly assume that the observed Vega spectrum has zero magnitudes in all filters. As demonstrated in the Appendix B of \citet{Rub15} (and earlier by \citealt{Rubele_etal12}), this definition implies the following corrections to bring the model Vega magnitudes onto the same system of the CASU v1.3 calibrations:
\begin{eqnarray}
Y_\mathrm{model}-Y_{1.3} &=& 0.074  \mathrm{~mag} \label{eqaa}\\
J_\mathrm{model}-J_{1.3} &=& 0.026  \mathrm{~mag}\\
\ks_\mathrm{model}-\ks_{1.3} &=& 0.003 \mathrm{~mag} \label{eqac}
\end{eqnarray}
Therefore, the right-hand numbers are subtracted from the isochrones, before the SFH work is done. The main effect of these correction is that of shifting the models to bluer values of \yks.

Corrections \ref{eqaa} and \ref{eqac} were adopted both in \citet{Rub15} and in the present analysis. Since they should no longer be relevant with the v1.5 recalibration (since our models and v1.5 data are expected to be in the same Vegamag system), the expected mean offset between our present analysis of v1.3 data, and the analysis of the new v1.5 data using the same stellar models, should be
\begin{eqnarray}
Y_\mathrm{model}-Y_{1.5} &=& 0.092  \mathrm{~mag}\\
J_\mathrm{model}-J_{1.5} &=& 0.006  \mathrm{~mag}\\
\ks_\mathrm{model}-\ks_{1.5} &=& 0.0136 \mathrm{~mag}
\end{eqnarray}
If we convert these differences into the equivalent values of colour excess, $E(\yks)$ and $E(\jks)$, and then into extinction differences, it turns out that with the new photometric v1.5 calibration our extinction values would be different by $\Delta\avyk=-0.29$~mag, and $\Delta\avjk=+0.045$~mag, on average.  While these systematic differences would explain why the \avyk\ values systematically larger than \avjk\ in some central areas of the SMC Bar and Wing (Fig.~\ref{fig.AVD}), they cannot explain the relatively large variation in $\avyk-\avjk$ from subregion to subregion, nor the subregions with a negative $\avyk-\avjk$, which are concentrated in the tiles SMC 3\_5 and 6\_3. Therefore, it is still unclear whether we can derive consistent values for \av\ by simply using the new calibration. In this paper, we prefer to present the derived extinction and distance values and discuss their general trends, without making too strong statements about their absolute values. We also prefer to discuss the extinction values derived from the $J\ks$ filters, for which the present results are expected to be more robust.

\label{lastpage}
\end{document}